\documentclass[pra,twocolumn,superscriptaddress,showpacs, amsmath,amstex,amssymb,citeautoscript]{revtex4-1}

\usepackage{bm}

\usepackage[T1]{fontenc}
\usepackage{natbib}
\usepackage[utf8]{inputenc}
\usepackage{lipsum}
\usepackage{amsmath}
\usepackage{amssymb}
\usepackage{bbm}
\usepackage{braket}
\usepackage{xcolor}
\usepackage{pifont}
\usepackage[frak=esstix]{mathalpha}
\usepackage[mathscr]{euscript}
\usepackage[shortlabels]{enumitem}
\usepackage[justification=justified]{subcaption}
\usepackage{graphicx}
\usepackage{lipsum}

\allowdisplaybreaks
\usepackage{float}
\usepackage{graphicx}
\usepackage{dsfont}
\usepackage{comment}
\usepackage[colorlinks=true]{hyperref}  
\hypersetup{
    bookmarks=true,         
    unicode=false,          
    pdftoolbar=true,        
    pdfmenubar=true,        
    pdffitwindow=false,     
    pdfstartview={FitH},    
    pdftitle={Spin-orbit crossover and the origin of magnetic torque in kagome metals},    
    pdfauthor={Scheurer and Scammell},     
    pdfsubject={},   
    pdfcreator={},   
    pdfproducer={}, 
    pdfkeywords={} {} {}, 
    pdfnewwindow=true,      
    colorlinks=true,       
    linkcolor=blue, 
    citecolor=blue,        
    filecolor=magenta,      
    urlcolor=blue           
} 
\usepackage{amsfonts}
\usepackage{upgreek}
\usepackage{slashed}
\usepackage{latexsym}

\newcommand{\beq}{\begin{equation}}
\newcommand{\eeq}{\end{equation}}
\def\bea{\begin{eqnarray}}
\def\eea{\end{eqnarray}}
\newcommand{\Tr}{\text{Tr}}

\usepackage{multirow}

\newcommand{\equref}[1]{Eq.~(\ref{#1})}

\newcommand{\pdagger}{{\phantom{\dagger}}}

\renewcommand{\approx}{\simeq}

\usepackage{braket}
\usepackage{comment}
\usepackage{slashed}
\usepackage{subcaption}

\renewcommand{\vec}[1]{\boldsymbol{#1}}

\usepackage{comment}
\usepackage{pifont}

\usepackage{relsize}

\usepackage{soul}

\usepackage{pgfplots}
\usepackage{ifthen}

\begin{document}

\title{Spin-orbit crossover and the origin of magnetic torque in kagome metals}

\author{Mathias S.~Scheurer}
\affiliation{Institute for Theoretical Physics III, University of Stuttgart, 70550 Stuttgart, Germany}

\author{Harley D.~Scammell}
\affiliation{School of Mathematical and Physical Sciences, University of Technology Sydney, Ultimo, NSW 2007, Australia}


\begin{abstract}
Recent experiments on the kagome metal CsV$_3$Sb$_5$ reveal a curious phase transition-like feature: a nematic magnetic torque response that abruptly sets in at $T_\tau \approx 130$~K, above the known charge density wave transition at $T_\text{CDW} \approx 100$~K. Counterintuitively, elastoresistance measurements—a standard probe of nematicity—show no corresponding signal, ruling out a nematic phase transition and placing strong constraints on possible explanations. Beyond nematicity, the torque is paramagnetic for in-plane magnetic field, while above a critical out-of-plane field, an in-plane magnetisation appears, accompanied by hysteresis.
We show that this combination of features cannot be accounted for by charge density waves or intraband magnetic order. Instead, we propose that interband ordering—via a symmetry-allowed interband spin-orbit coupling and a time-reversal and spatial symmetry-breaking interband order parameter—together with a background strain field, consistent with typical experimental conditions, provides a natural explanation; in our picture, the behaviour at $T_\tau$ is understood as a crossover in the symmetry-allowed interband spin-orbit coupling strength.
Our theory accounts for the nematic magnetic torque, hysteresis, and the transition-like onset at $T_\tau$, while also making testable predictions, including strain-induced magnetisation. In doing so, it challenges the prevailing view of the normal state.
\end{abstract}

\maketitle

\section{Introduction}
Kagome metals AV$_3$Sb$_5$,  ScV$_6$Sn$_6$ and related, have provided physicists a new playground to complex, intertwined phases of matter~\cite{wang2023quantum, yin2022topological, neupert2022review}. The familiar story prevails: a high-temperature charge-density wave (CDW), appearing at $T_\text{CDW}\sim100$K, gives way to a low-temperature superconductor ($T_c\sim2$K), with a host of puzzling phases in between~\cite{PhysRevMaterials.3.094407, PhysRevLett.125.247002, yang2021giant, Chen2022anomalous, Mielke2021b, jiang2021np, Xu2022, li2021rotation, saykin2023high, zhao2021cascade, Xu2021, Gupta2021, Duan2021, Li2021c, Shumiya2021, Miao2021, Ni2021, Zhu2021, Chen2021b, Du2021b, Zhang2021, Liang2021, ortiz2021fermi, oey2022fermi, kang2023charge, huang2025revealing, hossain2025field}.

Naturally, the CDW is the focus of most investigations, e.g. Refs.\cite{ParkPRB2021, christensen2021theory, WagnerPRB2023}—after all, it defines the effective normal state out of which all subsequent symmetry-breaking emerges. However, very recent experimental work on CsV$_3$Sb$_5$ \cite{Asaba2024} identifies a non-trivial response at $T_\tau\approx130$~K, i.e. above $T_\text{CDW}$. We argue that this response cannot be explained by CDW physics (e.g. pre-formed or fluctuating order), but instead signals overlooked aspects of the normal state as well as the ordering tendencies these precipitate. We therefore take the view that this feature is a vital piece of the kagome puzzle—since it supersedes the CDW, it arguably must be understood before one can propagate to lower temperatures.

We summarise the key experimental observations reported in Ref.~\cite{Asaba2024}.
First, a magnetic torque response is observed. In particular, the torque $\bm{\tau} = \bm{m} \times \bm{B}$, with $\bm m$ the magnetisation and $\bm B$ the applied field, exhibits a two-fold angular dependence: for an in-plane field $\bm{B} = B(\cos\phi, \sin\phi, 0)$, the torque component $\tau_z(\phi)$ behaves as $\tau_z \sin 2\phi$. Second, for in-plane fields, the response is paramagnetic. Above a certain critical out-of-plane field, the torque exhibits hysteresis as $\phi$ is swept, implying that an out-of-plane magnetic field induces in-plane magnetisation. Third, the torque response sharply kicks in at $T_\tau\approx130$K and grows approximately linearly, until reaching a plateau, as temperature is reduced. Noteworthy is that the response sets in above the known CDW transition $T_\text{CDW}\approx100$~K, and smoothly persists into the CDW phase. Moreover, the onset scale $T_\tau$ is independent of the applied magnetic field strength, indicating that there is a transition-like behaviour of the normal state at this scale.
Finally, elastoresistance measurements performed at $\bm{B} = \bm{0}$ detect no response in the $E_{2g}$ symmetry channel near $T_\tau$. This rules out spontaneous $C_3$ breaking at $T_\tau$. More generally, to our knowledge, no other symmetry-breaking transition near $T_\tau$ has been reported in the literature. 

This combination of observations and hence constraints is quite puzzling. In this work, we claim that {\it intra}band ordering---including spin and orbital magnetism as well as fluctuating  CDWs---cannot satisfy these constraints. Instead, we propose a resolution centring on {\it inter}band ordering. The key kinematic object we propose is an interband spin-orbit coupling (SOC), which we demonstrate to be a symmetry-allowed contribution to the Hamiltonian, i.e. respecting the point group D$_{6h}$. Methodologically, we treat these as variational parameters, denoted $\mathtt{g}$, in the free energy expansion. Unlike $\mathtt{d}$, which are symmetry-breaking interband order parameters belonging to non-trivial spatial irreducible representations (IRs), the $\mathtt{g}$ are symmetry-preserving and encode allowed interband spin-orbit couplings. They are not order parameters, but their temperature dependence governs key responses. Finally, we allow for a strain field, and account that this is non-zero due to extrinsic stress from the substrate; strain fields are ubiquitous in kagome metals unless carefully isolated \cite{emcha_Guo2024}. 

With these ingredients, we explain the key properties listed above: the two-fold angular dependence of the torque $\tau_z(\phi)$; the paramagnetic response to in-plane fields; and the existence of an out-of-plane magnetic field-induced in-plane magnetisation. In our theory, this last response emerges as part of a more general piezomagnetic response. All such observations rely on $\mathtt{g}$. The final puzzling observation is the sharp onset at $T_\tau$, yet with no response in elastoresistance or any other indication of a phase transition. This rules out, e.g. spontaneous ordering of $\mathtt{d}$ (or any order parameter that breaks $C_3$). Instead, an elegant solution arises: it is the symmetry-preserving $\mathtt{g}$ that undergoes a {\it crossover} at the scale $T_\tau$—mimicking a phase transition within the Landau formalism, yet with the crucial distinction that $\mathtt{g}$ preserves all symmetries of the Hamiltonian. Tuning the crossover scale to $T_\tau$, direct microscopic calculation produces a temperature dependence in agreement with observation—i.e. near-linear scaling, coming to a plateau as $T$ is lowered.

\section{Model}
\subsection{Hamiltonian}
Our starting point is a tight-binding Hamiltonian comprising $d_{xz}$ and $d_{z^2}$ orbitals on a kagome lattice with point group $D_{6h}$ (see Appendix \ref{A:TB} for details). A van Hove singularity (vHS) coming from each orbital lies in the vicinity of the Fermi energy \cite{Kang2022,Hu2022,Thomale_PRL2021,Scammell2023}. To capture this feature, we construct the low-energy effective two-band model
\begin{align}
    H_0 = \sum_{\bm k} \left( \varepsilon_{\bm k}^c c_{\bm k}^\dag  c^\pdagger_{\bm k} + \varepsilon_{\bm k}^v v_{\bm k}^\dag  v^\pdagger_{\bm k} \right), \label{H0Form}
\end{align}
where $c_{\bm k}^\dag$ ($v_{\bm k}^\dag$) are the creation operators for conduction (valence) electrons and $\varepsilon_{\bm k}^c$ ($\varepsilon_{\bm k}^v$) are the associated band energies. Throughout, we work in this orbital basis. We will use the term {\it interband} to refer to couplings between the two low-energy orbitals near the Fermi level.
In $H_0$ we have neglected interband couplings, which we will account for perturbatively below. These will be classified by their transformation under irreducible representations (IRs) of the point group. Later we will refer to spectral particle-hole symmetry (PHS), this corresponds to $\varepsilon_{\bm k}^v=-\varepsilon_{\bm k}^c$.

{\bf Interband coupling---}Crucially, the product of the IRs of the two bands is $A_{2u}$ of $D_{6h}$ [see Appendix \ref{A:TB}], i.e.~odd under inversion but trivial under six-fold rotational symmetry $C_{6z}$ and the reflections $\sigma_v$, $\sigma_d$. 
More precisely, we can choose a gauge of the Bloch states such that time-reversal, with anti-unitary Fock-space operator $\Theta$, acts as $\Theta c_{\vec{k}} \Theta^\dagger = i s_y c_{-\vec{k}} $, $\Theta v_{\vec{k}} \Theta^\dagger = i s_y v_{-\vec{k}}$, and the composite operator $F_{\bm k} = c_{\bm k}^\dag v_{\bm k}$ behaves as $F_{\bm k} \rightarrow F_{g\bm k}$ under the point symmetries $g\in \{C_{6z}, \sigma_v, \sigma_d\} $; meanwhile, under inversion $F_{\bm k} \rightarrow - F_{\bm k}$.
Despite the point group being inversion symmetric, this odd-parity interband overlap allows for a type of {\it interband spin-orbit-coupling}.

Then inversion and time-reversal symmetry demand that the interband Hamiltonian be of the form
\begin{equation}
    H_{\text{inter}} = \sum_{\bm k} c_{\bm k}^\dag\left( i g_{0,\bm k} +  \bm g_{\bm k}\cdot \bm s\right) v_{\bm k} + \text{H.c.}, \label{interbandCoupling}
\end{equation}
with $g_{0,\bm k} = - g_{0,-\bm k}$ and $\bm g_{\bm k} = -\bm g_{-\bm k}$ real-valued and Brillouin-zone periodic functions; the first term corresponds to a spin-independent coupling between the two bands which we have separated from $H_0$ \eqref{H0Form} since then the conduction and valence bands of $H_0$ correspond to the orbital basis, and $g_{0,\bm k}$ is then understood as the mixing of orbitals. Ultimately, this $g_{0,\bm k}$ will not play a role in the mechanism considered here, and one could envisage absorbing $g_{0,\bm k}$ into a redefinition of the bands. 
The second term in \equref{interbandCoupling} is the aforementioned inversion-symmetric spin-orbit coupling that is induced by the atomic spin-orbit interaction. Since $c_{\bm k}^\dag v_{\bm k}$ transforms under $A_{2u}$, the remaining symmetries of $D_{6h}$ require that the functions $g_0$, $(g_{x}, g_y)$, and $g_z$ transform under $A_{2u}$, $E_{1u}$, and $A_{1u}$, respectively. For convenience and later reference, we introduce the four-component vector $g_{\mu,\vec{k}} = (g_{0,\bm k},\vec{g}_{\bm k})_\mu$, with $\mu=0,1,2,3$, and write
\begin{align}
g_{\mu, \bm k} &= \mathtt{g}_{\mu} \chi_{\mu}(\bm k), \label{gExpansion}
\end{align}
where $\mathtt{g}_\mu \in \mathbb{R}$ are formal expansion coefficients used in our Ginzburg–Landau theory to organise the symmetry-allowed interband couplings in \equref{interbandCoupling}. The functions $\chi_\mu(\bm k)$ transform under the following irreducible representations:
\[
\begin{array}{c|c|c}
\mu & \text{IR of } D_{6h} & \chi_\mu(\bm k) \\ \hline
0 & A_{2u} & Z(\bm k) \\
1 & E_{1u} & Y(\bm k) \\
2 & E_{1u} & -X(\bm k) \\
3 & A_{1u} & Z(\bm k) R^z(\bm k) \\
\end{array}
\]
Here, $X$, $Y$, and $Z$ transform as $x$, $y$, and $z$ under $D_{6h}$, and $R^z(\bm k)$ as $xy(x^2 - 3y^2)(3x^2 - y^2)$. Defining $\psi^\dag_{\bm k} = (c^\dag_{\bm k}, v^\dag_{\bm k})^T$, with Pauli matrices $\sigma_j$ acting in this two-band space, and introducing the projectors $\sigma_c \equiv (\sigma_0 + \sigma_z)/2$ and $\sigma_v \equiv (\sigma_0 - \sigma_z)/2$, the total Hamiltonian $H = H_0 + H_{\text{soc}}$ reads
\begin{align}
\label{H}
H &= \sum_{\bm k} \psi^\dag_{\vec{k}} \left( \varepsilon_{\bm k}^c \sigma_c + \varepsilon_{\bm k}^v \sigma_v + g_{0,\bm k} \sigma_y + \bm g_{\bm k} \cdot \bm s \, \sigma_x \right) \psi^\pdagger_{\vec{k}}.
\end{align}
 
{\bf Symmetry breaking fields}---We 
include a second class of interband coupling, $d_{\mu,\bm k}$. As opposed to the $g_{\mu,\bm k}$ terms above, these are symmetry-breaking order parameter (OP) fields; breaking TRS and spatial symmetries. 
As we discuss in Appendix~\ref{IntrabandTerms}, neither (fluctuating) CDWs or intra-band magnetic order parameters are sufficient for explaining the magnetic torque experiments. We will therefore focus on translationally invariant, interband  terms of the form
\begin{equation}
    H_\text{OP}= \sum_{\bm k} \psi_{\bm k}^\dag (d_{0,\bm k} \sigma_x +  \bm d_{\bm k}\cdot \bm s \sigma_y)\psi^\pdagger_{\bm k}, \label{HOPExpr}
\end{equation}
where $d_{0,\vec{k}} = -d_{0,-\vec{k}}$ and $\vec{d}_{\vec{k}} = -\vec{d}_{-\vec{k}}$ as dictated by time-reversal symmetry. The OPs $d_{\mu,\bm k}$ are interaction induced, and will thus generically break symmetries of $D_{6h}$. Therefore, we will expand    
\begin{align}
    d_{\mu, \bm k}&=\sum_\Gamma \sum_{n=1}^{d_\Gamma} \mathtt{d}_{\mu,(\Gamma,n)}\Xi_{\Gamma,n}(\bm k), \label{dExpansion}
\end{align}
where $\Xi_{\Gamma,n}(\bm k)$ is the $n$th basis function transforming under IRs $\Gamma$ of $D_{6h}$ and $\mathtt{d}_{\mu,(\Gamma,n)}$ are real-valued expansion coefficients, similar to $\mathtt{g}_{\mu}$ in \equref{gExpansion}. As mentioned above, $d_{\mu, \bm k}$ have to be odd in $\vec{k}$ such that we restrict the sum over $\Gamma$ in \equref{dExpansion} to even IRs. Note: While $\chi_{\mu}(\bm k)$ and $\Xi_{\Gamma,n}(\bm k)$ both represent IR-respecting basis functions, the former denotes a fixed subset relevant to interband SOC, while the latter indexes general basis functions used in the order parameter expansion.

In principle, we could further have terms like
\begin{equation}
    \widetilde{H}_\text{OP}= \sum_{\bm k} \psi_{\bm k}^\dag (\tilde{d}_{0,\bm k} \sigma_y +  \tilde{\bm d}_{\bm k}\cdot \bm s \sigma_x)\psi^\pdagger_{\bm k}. \label{OtherOrderParameters}
\end{equation}
In this case, however, we have $\tilde{d}_{0,\vec{k}} = \tilde{d}_{0,-\vec{k}}$ and $\tilde{\vec{d}}_{\vec{k}} = \tilde{\vec{d}}_{-\vec{k}}$ such that they transform under even IRs of $D_{6h}$. As we will see below, this is why they cannot contribute in Zeeman coupling [\equref{gbdCoupling}] and so they will not play any role in our analysis.

Finally, we include magnetic and strain fields, $\bm B$ and $\bm\varepsilon$, with corresponding Hamiltonian contributions
\begin{align}
\label{HB}
H_{B}&= \sum_{\bm k} \psi_{\bm k}^\dag \bm B\cdot \bm s \sigma_0\psi^\pdagger_{\bm k},\\
\label{Hstrain}
H_\varepsilon &= \sum_{\bm k} \psi_{\bm k}^\dag (\bm \varepsilon \cdot (\Xi_{E_{2g},1}(\bm k),\Xi_{E_{2g},2}(\bm k)) \sigma_0)\psi^\pdagger_{\bm k}.
\end{align}
We have again made use of the basis functions $\Xi_{\Gamma,n}(\bm k)$ introduced in \equref{dExpansion}. Note that although the form of the basis function entering $d_{\mu,\vec{k}}$ and the strain coupling $H_\varepsilon$ does not have to be identical, $\Gamma = E_{2g}$ does not enter in \equref{HOPExpr} as it is even; as such, there is no need to introduce a separate symbol in \equref{Hstrain} for the basis functions.

\subsection{Free energy}
From the full MF Hamiltonian, given by $H_\text{MF} = H + H_\text{OP} + H_B + H_\varepsilon$, we obtain the free energy via the standard procedure of integrating out fermions and expanding in powers of $\mathtt{d}_{\mu,(\Gamma,n)}$, $\mathtt{g}_{\mu}$ as well as magnetic and strain fields $\bm B, \bm \varepsilon$. We denote by
\begin{align}
\label{F_full}
    {\cal F}[\mathtt{g},\mathtt{d},B,\varepsilon]&={\cal F}_0[\mathtt{g},\mathtt{d}] + {\cal F}_B[\mathtt{g},\mathtt{d},B] + {\cal F}_{\varepsilon}[\mathtt{g},\mathtt{d},\varepsilon]
\end{align}
the full free energy,
with each contribution representing, respectively, the zero-field contribution, the magnetic field contribution, and the strain contribution. 

{\bf Zero-field---}${\cal F}_0$ is a polynomial in $\mathtt{d}_{\mu,(\Gamma,n)}$, $\mathtt{g}_{\mu}$, the lengthy expression is relegated to the SM. We do make a special note that, based on the experimental findings, we are treating the quadratic coefficients with ${\cal F}_0$ as positive; we do not consider spontaneous symmetry breaking in ${\cal F}_0$. Instead, for the temperatures of interest $T\approx T_\tau$ the $\mathtt{d}_{\mu,(\Gamma,n)}$ can acquire expectation values only upon application of magnetic field (with or without strain). 

{\bf Zeeman---}Considering first the magnetic field, and expanding to linear order, we find a nonzero coupling of $\mathtt{d}_{\mu,(\Gamma,n)}$ to the Zeeman field $\vec{B}$, which by symmetry must involve the interband terms $\mathtt{g}_{\mu}$. Upon specialising to spectral particle-hole symmetry (PHS), this coupling reduces to, schematically, $\bm B\cdot (\mathtt{d}\times\mathtt{g})$. Relaxing PHS, one would find additional contributions of the form $\mathtt{d}_0 \mathtt{g} \cdot \vec{B}$. Explicitly, we obtain
\begin{align}
{\cal F}_B
    &= \sum_{i,j,k = 1}^3 \sum_{\Gamma,n} c_i B_i \mathtt{d}_{j,(\Gamma,n)} \mathtt{g}_{k} \varepsilon_{ijk}\delta_{\Gamma,\Gamma_k}\delta_{n,n_k}. \label{gbdCoupling}
\end{align}
Here, $c_i$ are real constants obeying $c_x=c_y\neq c_z$, which follow from loop integral calculations; expressions are provided in the SM. To clarify, here $\Gamma_k = (A_{2u},E_{1u},A_{1u})_k$ and hence $n_{k} = (1,2,1)_{k}$. The Kronecker deltas $\delta_{\Gamma,\Gamma_k}\delta_{n,n_k}$ just follow from orthogonality of different basis functions. It follows that the order parameters in \equref{OtherOrderParameters}, which transform under even IRs as opposed to the components of $g_{\mu,\vec{k}}$, therefore do not exhibit a Zeeman coupling, i.e. they do not contribute to \equref{gbdCoupling}.  
By the same reasoning, \equref{gbdCoupling} does not contain all odd IRs, but instead only a subset of terms in \equref{dExpansion} can contribute; this subset is dictated by the specific IRs of $g_{\mu,\vec{k}}$. To represent them more compactly in the following, we write $\mathtt{d}_{j,x} := \mathtt{d}_{j,(E_{1u},1)}$, $\mathtt{d}_{j,y} := \mathtt{d}_{j,(E_{1u},2)}$, and $\mathtt{d}_{j,z} := \mathtt{d}_{j,(A_{1u},1)}$. Due to anti-symmetry of the $\varepsilon$-tensor in \equref{gbdCoupling}, only six distinct terms enter and we collect them into the vector 
\begin{align}
\label{d_6D}
\bm{\mathtt{d}} &= 
\begin{pmatrix} 
    \mathtt{d}_{1,x}, & \mathtt{d}_{1,z}, & 
    \mathtt{d}_{2,y}, & \mathtt{d}_{2,z}, & 
    \mathtt{d}_{3,x}, & \mathtt{d}_{3,y}
\end{pmatrix} \in {\mathbb R}^6.
\end{align}

The upshot: Experimentally observes a paramagnetic response. By considering interband order $\mathtt{d}$, and invoking the interband spin-orbit coupling $\mathtt{g}$, we construct a Zeeman coupling between $\bm B$ and $\mathtt{d}$ and hence paramagnetic response. Crucially, this coupling would be forbidden without $\mathtt{g}$ and, instead, there would be a threshold $B_c$, above which a magnetisation is induced, below which magnetisation is zero. This is inconsistent with experiment, and thereby provides direct support for the role of the proposed interband SOC. 

{\bf Strain---}To explain the key features: (i) two-fold angular dependence of the torque; (ii) the hysteresis observed for $B_z$ beyond a critical strength, it will prove crucial to include the influence of a strain field $\bm\varepsilon=(\varepsilon_{x^2-y^2},\varepsilon_{xy})$. For clarity, we specialise to $\varepsilon_{x^2-y^2}$. 

There are two contributions, denoted ${\cal F}_\varepsilon={\cal F}^{(1)}_\varepsilon+{\cal F}^{(2)}_\varepsilon$. First, considering the lowest order coupling between a strain field $\varepsilon_{x^2-y^2}$ and the order parameters $\mathtt{d}_{\mu,(\Gamma,n)}$, we arrive at
\begin{align}
\label{Fstrain1}
 \notag &{\cal F}^{(1)}_\varepsilon=\varepsilon_{x^2-y^2}\Big\{b_0(
 \mathtt{d}_{1x}^2+\mathtt{d}_{3x}^2-\mathtt{d}_{2y}^2-\mathtt{d}_{3y}^2)\\
 \notag &+b_1(\mathtt{g}_{2} \mathtt{d}_{3x},-\mathtt{g}_{1} \mathtt{d}_{3y})\cdot (\mathtt{g}_{2} \mathtt{d}_{3x}+\mathtt{g}_{3} \mathtt{d}_{2z}, \mathtt{g}_{1} \mathtt{d}_{3y}-\mathtt{g}_{3} \mathtt{d}_{1z})\\
&+b_2(\mathtt{g}_{2} \mathtt{d}_{1x}-\mathtt{g}_{1} \mathtt{d}_{2y})(\mathtt{g}_2 \mathtt{d}_{1x}+\mathtt{g}_1 \mathtt{d}_{2y})
 \Big\}.
\end{align}
This form follows from $H_\text{MF}$, at $B=0$, upon integrating out the fermionic degrees of freedom.

\begin{figure*}[t!]  
\includegraphics[width=0.7\linewidth]{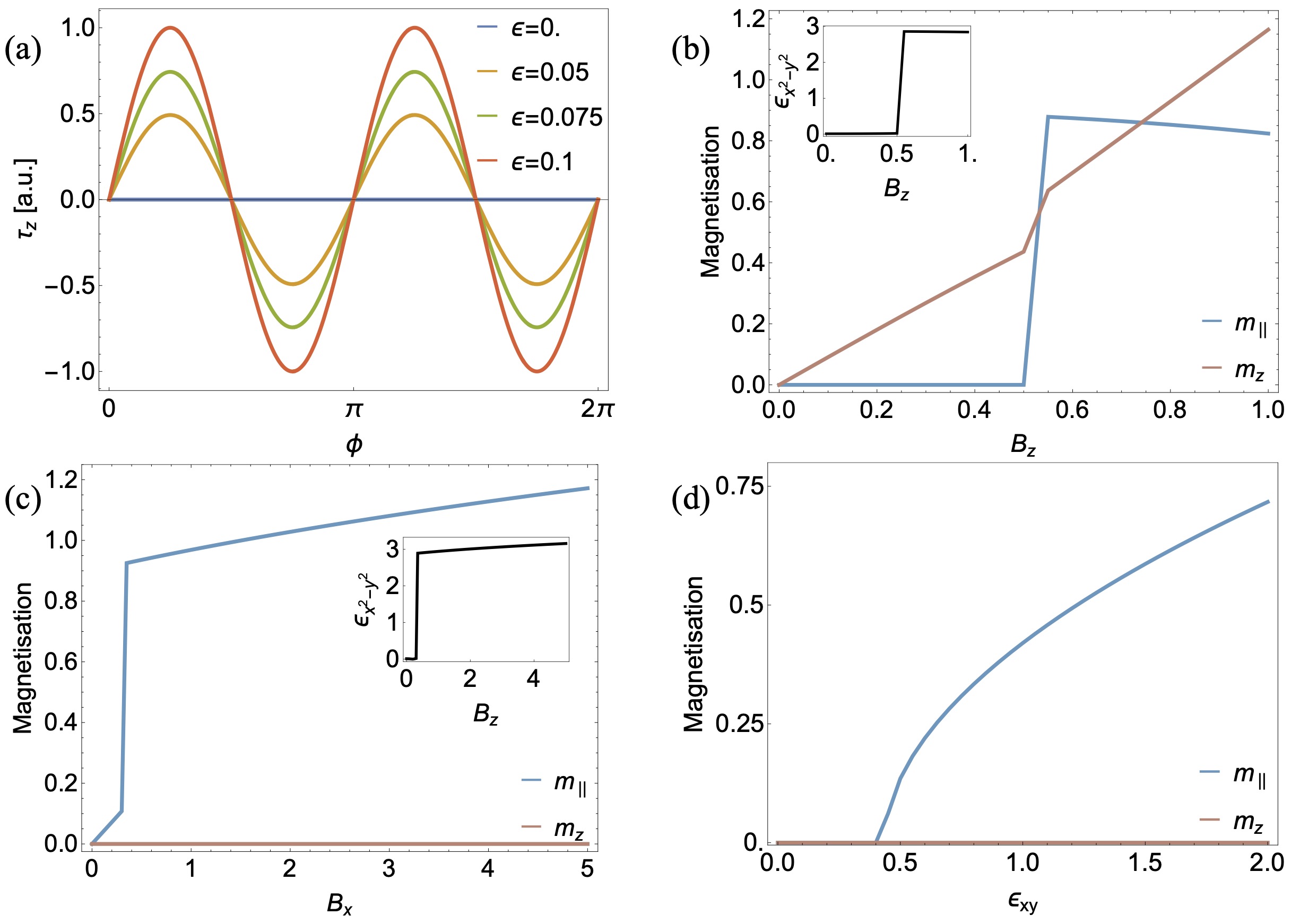}
    \caption{(A). Angular dependence of $\tau_z$, for purely in-plane field $\bm B = B (\cos\phi, \sin\phi,0)$, parametrised by strain $\varepsilon$. Curves are computed using the full model, ${\cal F}[\mathtt{g},\mathtt{d}, B, \varepsilon]$ in \equref{F_full}. (B) Magnetisation vs $B_z$, with (inset) $\varepsilon_{x^2-y^2}$ vs $B_z$. (C) Magnetisation vs $B_x$, with (inset) $\varepsilon_{x^2-y^2}$ vs $B_x$. (D) Magnetisation vs $\varepsilon_{x^2-y^2}$.  Everywhere the free energy parameters are: $\{\mathtt{g}_0, \mathtt{g}_1,\mathtt{g}_2,\mathtt{g}_3\} = \{0.1,0.5,0.5,0.75\}$, while quadratic and quartic coefficients are taken to be $\chi^{(2)}=1/16$ and $\chi^{(4)}=\left[\chi^{(2)}\right]^2$.}
    \label{fig:all}
\end{figure*}

Next, to account for the back-action of the order parameters $\mathtt{d}_{\mu,(\Gamma,n)}$ on the strain, one must promote $\varepsilon_{xy}$ to a field variable, with equilibrium value obtained from minimization of the free energy. The corresponding bare free-energy contribution is
\begin{align}
\label{Fstrain2}
    {\cal F}^{(2)}_\varepsilon&= -a_\varepsilon \varepsilon_{x^2-y^2} + \frac{1}{2}b_\varepsilon  \varepsilon_{x^2-y^2}^2.
\end{align}
Here  $a_\varepsilon>0$ accounts for extrinsic stress on the system, e.g. due to substrate, while $\varepsilon_{x^2-y^2}$ is the strain field of the system. Here $b_\varepsilon>0$ represents the lattice stiffness and is required for stability.  

Elastoresistance measurements, taken at $B=0$, do not show a response at $T_\tau$, implying there is no spontaneous $C_3$-breaking near $T_\tau$ (for $B=0$) nor is there any dramatic change in the $E_{2g}$ channel. To account for this observation, we model a small strain field, $\varepsilon_{x^2-y^2}$, which can be treated as approximately constant with temperature across $T_\tau$, i.e. $a_\varepsilon>0$ is assumed to be a small constant. The existence of the strain itself is consistent with a number of experiments on kagome metals showing that, unless carefully isolated \cite{emcha_Guo2024}, the lattice suffers from strain, likely due to the substrate.

\section{Results}
\subsection{Isotropic torque}
To guide the discussion, we first compute the torque without taking into account the strain field. This corresponds to taking ${\cal F}_{\varepsilon} = 0$ in \equref{F_full}. 
The Zeeman coupling induces the magnetisation, 
\begin{align}
\label{mag}
m_i &= -\partial_{B_i} {\cal F}^\text{min}_B
   &= c_x\begin{pmatrix} -\mathtt{g}_2 \mathtt{d}^0_{3x}-\mathtt{g}_3 \mathtt{d}^0_{2z} \\ \mathtt{g}_3 \mathtt{d}^0_{1z}-\mathtt{g}_1 \mathtt{d}^0_{3y}\\ \frac{c_z}{c_x}(\mathtt{g}_1 \mathtt{d}^0_{2y}+\mathtt{g}_2 \mathtt{d}^0_{1x}) \end{pmatrix}_i
\end{align}
here it is understood that the fields $\mathtt{d}^0_{i,\mu}$ are taken at the minima of the free energy at $\bm B\neq\bm0$; these $\mathtt{d}^0_{i,\mu}=\mathtt{d}^0_{i,\mu}(\bm B)$ depend on magnetic field. 
Evaluating the magnetisation, one finds that $(m_x,m_y)$ and $(B_x,B_y)$ are collinear, which implies that $\tau_z=(\bm m\times \bm B)\cdot \hat{z}=0$. To linear order in magnetic field, this readily follows from the underlying $D_{6h}$ symmetry; in our current model, however, it also holds to all orders in magnetic field. For later reference, an equivalent statement is that $\bm m = \hat{\chi}(\bm B)\cdot \bm B$, with $\chi_{xx}(\bm B)=\chi_{yy}(\bm B)$ and $\chi_{xy}(\bm B)=\chi_{yx}(\bm B) = 0$.

\subsection{Nematic torque}
Upon including the strain field, i.e. the contribution ${\cal F}_\epsilon$, an evaluation of the magnetisation and magnetic susceptibility of ${\cal F}$ \eqref{F_full} reveals that the in-plane rotational symmetry is broken; $\chi_{xx}\neq \chi_{yy}$. It follows that this produces nonzero magnetic torque $\tau_z$. Performing explicit minimsation of ${\cal F}$ \eqref{F_full}, and subsequent evaluation of $\chi_{ij}$, for a chosen set of parameters, produces the magnetic torque $\tau_z$ as shown in Fig. \ref{fig:all}, which exhibits the desired two-fold angular dependence as seen experimentally. 

To better understand the mechanism of two-fold angular dependent torque, we here appeal to an effective free energy, written in terms of the magnetisation $\bm m$ and in-plane $\bm B$, 
\begin{align}
    {\cal F}_\text{eff}&= - \bm m\cdot \bm B + a \bm m\cdot \bm m - \varepsilon (m_x^2-m_y^2).
\end{align}
This model is used purely for demonstration, all numerical results are obtained from ${\cal F}$ of \eqref{F_full}. 

Computing the magnetic susceptibilities $\chi_{xx},\chi_{yy}$  of ${\cal F}_\text{eff}$ and parametrising $\bm B = B (\cos\phi, \sin\phi,0)$, one finds non-zero torque $\tau_z$ with two-fold angular dependence, 
\begin{align}
    \tau_z &= \frac{1}{2}B^2(\chi_{xx}-\chi_{yy}) \sin2\phi = \frac{1}{2}\frac{\varepsilon}{a^2-\varepsilon^2} B^2\sin2\phi.
\end{align}
We also see how the magnitude of $\varepsilon$ influences the magnitude of $\tau_z$, clarifying the full model calculations presented in Fig. \ref{fig:all}.

\begin{figure*}[t!]  
\includegraphics[width=0.35\linewidth]{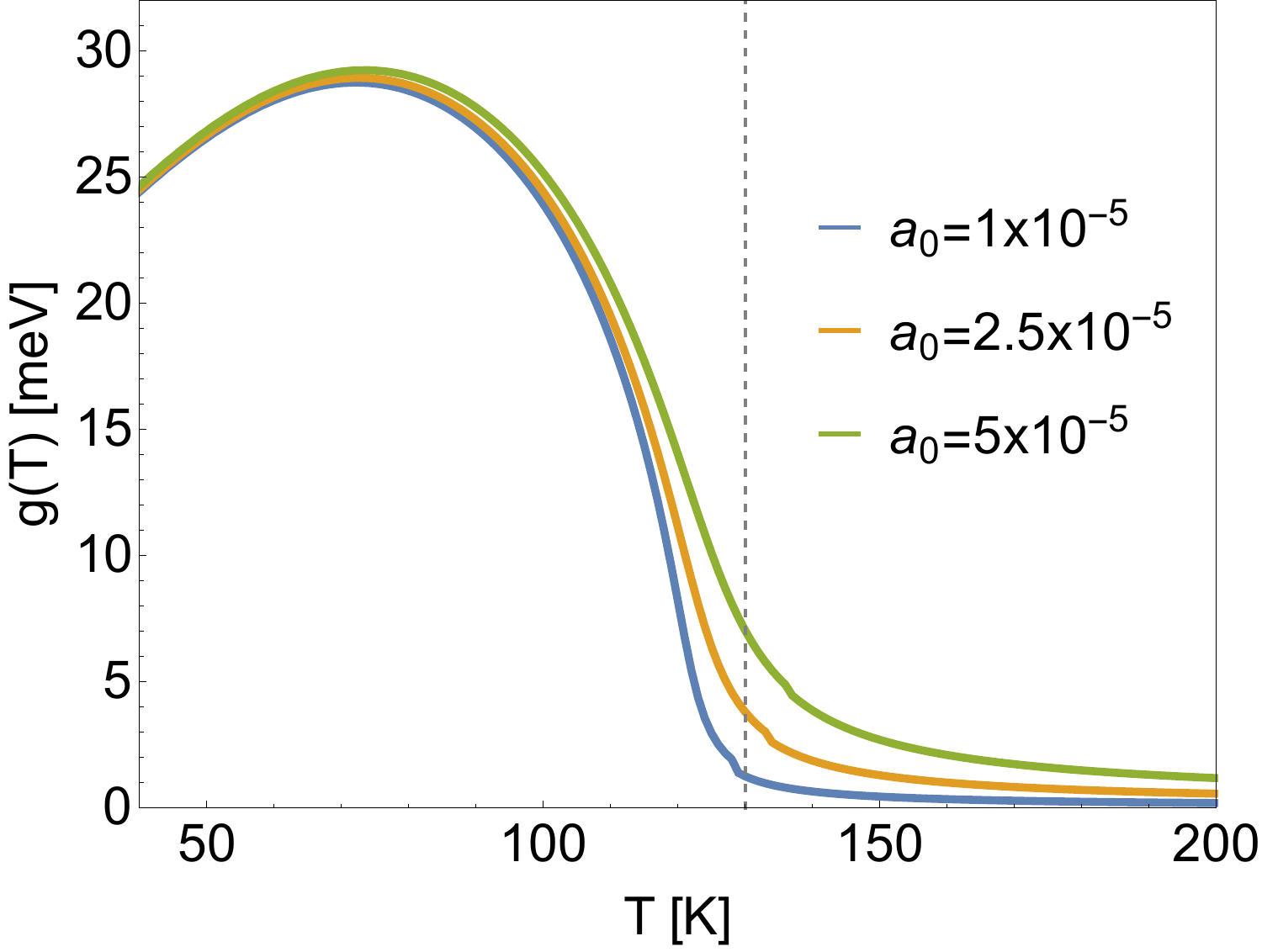}\hspace{0.5cm}
\includegraphics[width=0.35\linewidth]{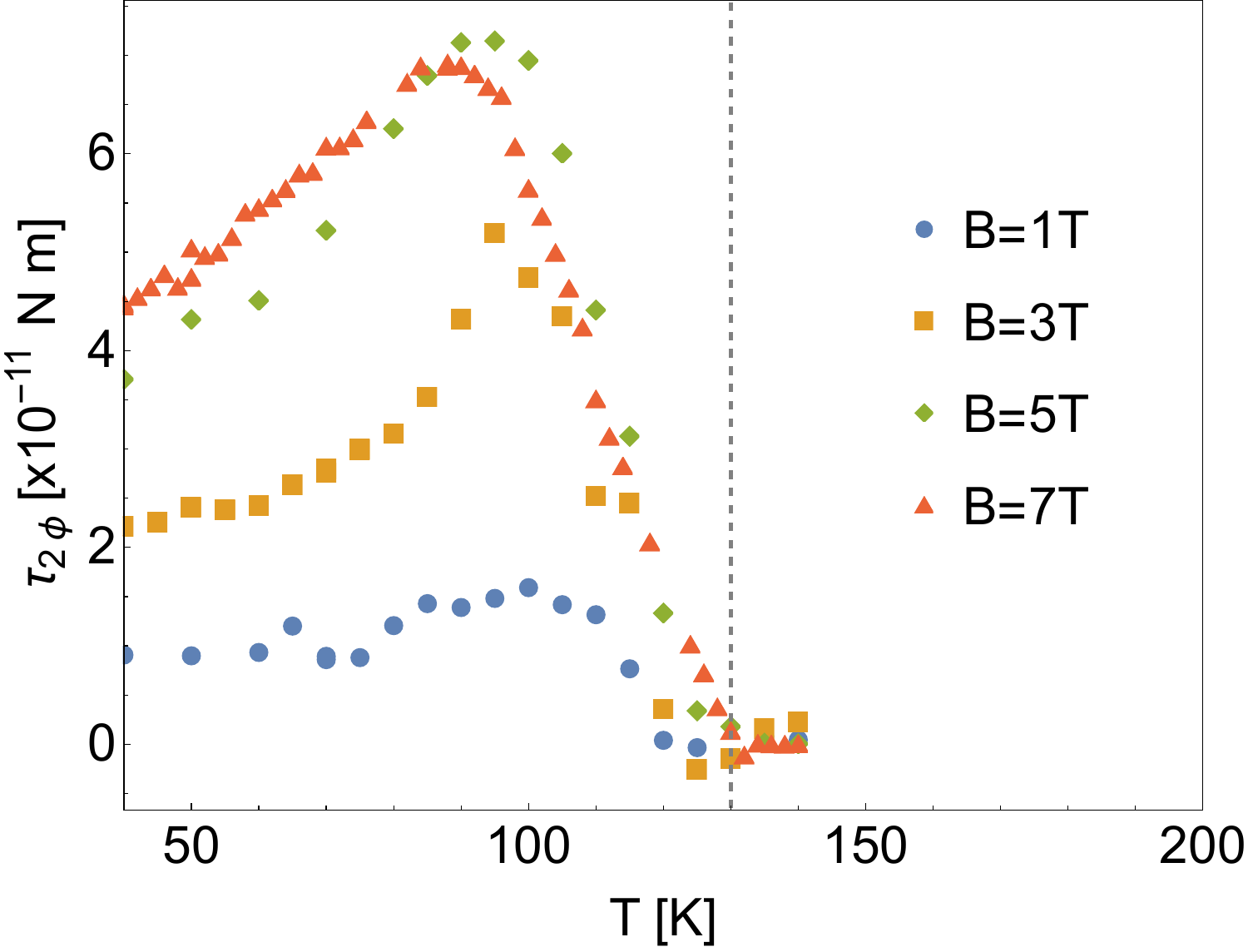}
    \caption{{\bf Crossover at $T_\tau$.} (a) Calculated $\mathtt{g}(T)$, for different values of $a_0$, from \eqref{FgToy}. Dashed vertical line indicates $T_\tau$. (b) Experimentally measured $\tau_{2\phi}$ in  $\tau_z = \tau_{2\phi} \sin2\phi$ at various magnetic field strengths; based on raw data from Asaba, T., Onishi, A., Kageyama, Y. et al available at \cite{Asaba2024}.}
    \label{fig:gonset}
\end{figure*}

\subsection{Hysteresis induced by $B_z$}
A prominent feature of the magnetic torque experiment is the occurrence of a hysteresis for $B_z>B_z^c$. This translates to: for sufficiently strong $B_z$, an in-plane magnetisation is induced. Indeed our modeling ${\cal F}$ contains such a feature. Explicitly, the contribution ${\cal F}^{(1)}_\varepsilon$ contains coupling between strain and the same combination of $\mathtt{d}_{\mu,(\Gamma,n)}$ and $\mathtt{g}_{\mu}$  as the magnetisation \eqref{mag}.  Comparing the magnetisation components of Eq.~\eqref{mag} with the expression for strain coupling Eq. \eqref{Fstrain1}, we see that strain couples to $(m_x, m_y)$, via the $b_1$ term in \eqref{Fstrain1}, while it couples to $m_z$ via the $b_2$ coupling. Hence, for an appropriate choice of parameters, $B_z$ (and hence $m_z$) will enhance the strain field, and for sufficiently strong strain an in-plane $(m_x,m_y)$ can be induced. Minimization of the full free energy, subject to a certain choice of parameters, quoted in Fig. \ref{fig:all} caption, yields the desired $B_z$-induced in-plane magnetisation $m_\parallel$, see Fig. \ref{fig:all}(b). Subject to the same parameters, the in-plane magnetic field does not induce an out-of-plane magnetisation Fig. \ref{fig:all}(c).

This coupling can be understood as part of a broader, piezomagnetic response.  We tease out further implications of the piezomagnetic response below.

\subsection{Prediction: Piezomagnetism}

Beyond explaining existing observations, our model makes a concrete prediction: the presence of a piezomagnetic response. Notably, such a response has already been reported in RbV$_3$Sb$_5$ using laser-coupled scanning tunnelling microscopy \cite{Xing2024}.

As shown in Fig.~\ref{fig:all}(b) and (c), the strain field exhibits a first-order jump as a function of magnetic field. We therefore propose elastoresistance measurements as a function of $\bm B$ as a direct test of the predicted piezomagnetic coupling. Moreover, Fig.~\ref{fig:all}(d) also reveals that increasing in-plane strain beyond a critical threshold induces an in-plane magnetisation. This provides an additional experimental handle—strain-controlled magnetisation—as a further consequence of the same underlying response.

\subsection{Spin-orbit crossover at $T_\tau$}

Our modelling establishes the presence of an interband paramagnetic response which, when coupled to strain, accounts for the experimentally observed magnetic torque and hysteresis. However, a key puzzle remains: in experiment, the magnetic torque turns on rather abruptly below a characteristic temperature $T_\tau$. This is a puzzle since elastoresistance measurements across $T_\tau$ show no sign of a response, ruling out spontaneous nematicity. This imposes a significant constraint on possible explanations.

We propose a resolution in terms of interaction-driven thermal renormalisation of the $\mathtt{g}_{\mu}$; this acts analogous to a phase transition, insofar as there is a dramatic enhancement of $\mathtt{g}_{\mu}$ at some temperature, yet with the important distinction that no symmetries are broken. We therefore refer to it as a crossover. By virtue of \equref{mag}, the crossover acts as a sudden enhancement of the magnetic $g$-factor, and hence of the torque response. 


Setting $B=0$, such that the $\mathtt{d}_{\mu, (\Gamma,n)}$ are gapped, non-critical fields, then we may restrict our attention to the low-energy $\mathtt{g}_{\mu}$-sector. Within this sector, we examine the possible crossover behaviour in $T$. The form of the free energy is, restricting to $\vec{\mathtt{g}}=(\mathtt{g}_1,\mathtt{g}_2)$ for clarity,
\begin{align}
\label{FgToy}
   \hspace{-0.25cm} {\cal F}[\mathtt{g}] &= -a_0(\mathtt{g}_1+\mathtt{g}_2) + \left[\frac{1}{V_\text{eff}} - a_2(T)\right]\vec{\mathtt{g}}\cdot\vec{\mathtt{g}} \\
   \notag &+(a_4(T)\mathtt{g}_1^4+2b_4(T)\mathtt{g}_1^2\mathtt{g}_2^2+a_4(T)\mathtt{g}_2^4).
\end{align}
Here $a_0\neq 0$ reflects the existence of symmetry allowed $\mathtt{g}\neq\bm 0$. The temperature dependence of the coefficients can be deduced to be $a_2(T)\sim \ln(\Lambda/T)$, $a_4(T),b_4(T)\sim \ln(\Lambda/T)/T^2$, see e.g. \cite{NandkishoreChubukovPRL2012}. In this work we compute them explicitly, numerically via
\begin{align}
  \notag  a_2(T) &= T\sum_n \int \frac{d^2k}{(2\pi)^2}\frac{X_{\bm k}^2}{(i\omega_n-\varepsilon^c_{\bm k})(i\omega_n-\varepsilon^v_{\bm k})} ,\\
  \notag  a_4(T) &= T\sum_n \int \frac{d^2k}{(2\pi)^2}\frac{X_{\bm k}^4}{(i\omega_n-\varepsilon^c_{\bm k})^2(i\omega_n-\varepsilon^v_{\bm k})^2},\\
   b_4(T) &= T\sum_n \int \frac{d^2k}{(2\pi)^2}\frac{X_{\bm k}^2Y_{\bm k}^2}{(i\omega_n-\varepsilon^c_{\bm k})^2(i\omega_n-\varepsilon^v_{\bm k})^2}, 
\end{align}  
with $\varepsilon^{c/v}_{\bm k}$ obtained from the tight binding model presented in Appendix \ref{A:TB}, taking hopping $t=1$~eV, chemical potential $\mu=1$~eV and on-site potential $\varepsilon_{\sigma=\pm}=\pm 7.5\times10^{-3}$~eV. 
 Note that generically $b_4(T)<a_4(T)$ and hence both components $\mathtt{g}_1,\mathtt{g}_2$ receive an enhancement, i.e.~their coexistence is energetically favoured. This property prevents spontaneous $C_3$ breaking due to enhancement of just one component.
Finally, the interaction enters via $V_\text{eff}$; taking $V_\text{eff}=1/a_2(T_\tau)$ sets the crossover scale to be $T=T_\tau$. This leaves $a_0$ as a free parameter; adjusting $a_0$, we arrive at Fig. \ref{fig:gonset}. It can be seen that our modelling nicely captures the crossover seen experimentally. Curiously, it also captures the existence of a maximum near $T=100$~K, which is right at the CDW transition.

\subsection{Alternate scenarios}
{\it CDW---}A key aspect of the derived free energy was the linear coupling of $\bm B$ to $\mathtt{d}_{\mu,(\Gamma,n)}$; as long as $\mathtt{d}_{\mu,(\Gamma,n)}$ are uncondensed, this coupling generates a linear-in-B magnetisation as per experiment. By contrast, if we suppose that multiple fluctuating, nearly condensed CDWs which break mirror and time-reversal symmetries are responsible for the torque measurements, then we find that the lowest coupling is $\bm B$ to a bilinear in CDWs. The only way to explain the torque is if all participating CDWs are critical. However, the torque measurements persist deep into the CDW phase, which by definition has nonzero CDW ordering. This is a contradiction and we thereby rule out CDWs as the mechanism. Detailed calculations are supplied in the SM. 

{\it Intraband orbital magnetism---}permits both a linear-in-$B$ magnetisation and a piezomagnetic coupling. However, direct calculation shows that the observed crossover behaviour does not follow [Appendix \ref{IntrabandTerms}]. One could imagine fine-tuning the system to lie near a Stoner instability at $T_s \lesssim T_\tau$, such that the response sharply increases near $T=T_\tau \gtrsim T_s$. However, upon lowering towards $T=T_s$, additional ingredients must be added so as to prevent the transition; since any true magnetic transition would have been detected, e.g.~via Kerr rotation. But this scenario, involving Stoner enhancement without a subsequent transition, does not emerge from mean-field theory; additional assumptions would be required to suppress the instability. By contrast, our proposed interband mechanism captures all key features within simple mean-field theory. We therefore regard the intra-orbital magnetism scenario as fine-tuned and unlikely.

{\it Intraband spin magnetism---}permits a linear-in-$B$ magnetisation, but without an {\it intraband} SOC, the observed two-fold angular dependence cannot be captured. Crucially, such SOC is forbidden by $D_{6h}$ symmetry, and we therefore rule out this scenario on symmetry grounds.

\

\section{Discussion}
We have introduced a symmetry-allowed, but previously overlooked, kinetic contribution to kagome metals: the interband spin-orbit coupling (SOC). This term is permitted despite the crystal preserving inversion symmetry and plays a crucial role in understanding recent experimental anomalies.

Our central aim was to reconcile the seemingly conflicting responses observed at $T \lesssim T_\tau$: namely, the emergence of a nematic magnetic torque, i.e. magnetic susceptibility with $\chi_{xx} - \chi_{yy} \neq 0$, despite the absence of a corresponding signal in elastoresistance; a puzzling in-plane magnetisation induced by an out-of-plane field, i.e. $\chi_{xz} \neq 0$; and crucially that there appears to be a sharp crossover in the system at $T_\tau$, with no associated symmetry breaking.

We emphasise that conventional explanations based on charge density waves or intraband magnetism were ruled out. Instead, a key insight is that the observed responses cannot be understood without considering both bands near the Fermi level. The inclusion of interband SOC is essential, as it enables a linear-in-$B$ behaviour of the magnetisation, consistent with experiment, and also provides a natural explanation of the non-symmetry breaking crossover observed at $T_\tau$. To account for the anomalous susceptibility components $\chi_{\mu\nu}$, we found it necessary to include a background strain field. We pointed out that strain fields are ubiquitous in kagome metals unless carefully isolated \cite{emcha_Guo2024}. Ultimately, the puzzling behaviour of $\chi_{\mu\nu}$ can be attributed to a piezomagnetic coupling. This perspective suggests a clear experimental direction: probing the joint response to magnetic field and in situ strain—for instance via piezoelectric control—could directly test the predicted piezomagnetic coupling and illuminate the role of the interband SOC crossover.

\paragraph*{Implications.} This work demonstrates that a proper account of the rich phase diagram of kagome metals requires incorporating both bands near the Fermi level, rather than modelling only one, as is commonly done. Including this additional orbital degree of freedom not only resolves the specific anomalies discussed here, but is also likely to reshape our understanding of the broader phase diagram—from charge density wave formation to the subsequent array of exotic charge orders, and perhaps even superconductivity.



%

\begin{appendix}

\begin{widetext}


\section{Tight Binding Model}\label{A:TB}

\subsection{Tight binding Hamiltonian} 
A tight binding model describing the situation, generic to the family AV$_3$Sb$_5$, of two oppositely dispersing van Hove bands near the Fermi level has been obtained in Ref. \cite{Thomale_PRL2021}. Specifically, they work with $xz$ and $yz$ orbitals, which is relevant to KV$_3$Sb$_5$. In this work we focus on CsV$_3$Sb$_5$, for which the relevant orbitals near the Fermi level are $xz$ and $z^2/x^2-y^2$, see e.g. Ref. \cite{Hu2022}. Following the structure of the tight binding model of Ref. \cite{Thomale_PRL2021}, or by symmetry, we write for CsV$_3$Sb$_5$,
\begin{align}
{\cal H}&={\cal H}_0 + {\cal H}_1,\\
{\cal H}_0&=-\sum_{\bm k, i j, \sigma}t_\sigma \Omega_{ij}(\bm k) \psi^\dag_{\bm kj\sigma}\psi^\pdagger_{\bm ki\sigma} + \sum_{\bm k, i, \sigma}\varepsilon_\sigma \psi^\dag_{\bm ki\sigma}\psi^\pdagger_{\bm ki\sigma},\\
{\cal H}_1&=-t'\sum_{\bm k, i j,\sigma}\Omega_{ij}(\bm k) s_{ij} g^\mu\sigma^\mu_{\sigma\bar\sigma}F_{j,\sigma}F_{i,\bar{\sigma}}\psi^\dag_{\bm kj\sigma}\psi^\pdagger_{\bm ki\bar{\sigma}}.
\end{align}
Here $i,j,k\in\{A,B,C\}$ enumerate sublattice, $\sigma$ enumerate orbitals, i.e. $\sigma\in\{d_{xz}, d_{z^2/x^2-y^2}\}=\{+,-\}$, and $\psi^\dag_{\bm k i \sigma}$ is electron creation operator. The $\Omega_{ij}(\bm k)=1+e^{-2i \bm k\cdot \bm a_{ij}}$ and the form factors $F_{i,\sigma}$ transform as $d_{xz}, d_{z^2/x^2-y^2}$, and are also indexed by site $i$. The combination $s_{ij} g^\mu\sigma^\mu_{\sigma\bar\sigma}$ (with $\mu=x,y$) transforms identically to $F_{j,\sigma}F_{i,\bar{\sigma}}$ under the point group symmetries, and hence the combination $s_{ij;\sigma\bar\sigma}F_{j,\sigma}F_{i,\bar{\sigma}}$ is trivial. Here the $g^\mu$ are dimensionless numbers, which ultimately give rise to the interband spin-orbit coupling used in the main text. We suggest that they can be obtained from {\it ab initio}. In this construction, since we extract the form factors, $F_{i,\pm}$, the remnant electronic operator, $c_{\bm ki\sigma}$, transforms trivially under all $D_{6h}$ operations.

\subsection{Relative band IRs with respect to $D_{2h}$}
Considering the band IRs at a given $M$-point, the little group is $D_{2h}$.  It has been established in, e.g.  Ref. \cite{Hu2022}, that the IRs of the two vHS bands of interest in terms of the little group $D_{2h}$ are: the  $p$-type vHS with $d_{z^2/x^2-y^2}$-orbital belongs to $A_g$  (inversion-even), while the $m$-type vHS with $d_{x z}$-orbital belongs to $\mathrm{B}_{1u}$ (inversion-odd). The product is simply $A_{g}\otimes B_{1u} = B_{1u}$, which transforms as $z$. 

Considering these states defined everywhere in the Brillouin zone, we may classify their IRs with respect to $D_{6h}$. Since $D_{6h}$ can be generated by extending $D_{2h}$ with $C_{3z}$ rotations, we focus on how the band IRs behave under $C_{3z}$. Finding that the band IRs transform trivially under $C_{3z}$, see Section \ref{sec:C3}, it follows that $B_{1u}$ of $D_{2h}$ goes to $A_{2u}$ of $D_{6h}$. 

\subsection{Transformation under $C_{3z}$}\label{sec:C3}
The effective Hamiltonian \equref{H0Form} of the main text is the band-projected version of ${\cal H}_0$. Using short-hand notation $f_{\vec{k}\sigma} = (c_{\vec{k}},v_{\vec{k}})_\sigma$, we have
\begin{equation}
    f_{\vec{k}\sigma} = \sum_{j} (\phi_{\vec{k}\sigma})_j  \psi_{\bm kj\sigma},
\end{equation}
where $\phi_{\vec{k}\sigma}$ are the associated (three-component) Bloch states. Let us now establish the transformation properties under $C_{3z}$. 

The eigenstates of the van Hove bands (at M-points) exhibit a {\it momentum-sublattice locking}. Denoting $\sigma_\pm$ as the $p$- and $m$-type vHS, respectively, the $p$-type eigenstates are
\begin{align}
\label{phi_ptype}
    \phi_{\bm M_1,\sigma_+}&=\hat{A},\quad \phi_{\bm M_2,\sigma_+}=\hat{B},\quad
    \phi_{\bm M_3,\sigma_+}=\hat{C}, 
\end{align}
where $\hat{A}, \hat{B}, \hat{C}$ are basis vectors ``along'' the three sublattices.
Meanwhile for $m$-type vHS, we have 
\begin{align}
\label{phi_mtype}
    \phi_{\bm M_1,\sigma_-}&=\frac{1}{\sqrt{2}}(\hat{B}+\hat{C}), \quad\phi_{\bm M_2,\sigma_-}=\frac{1}{\sqrt{2}}(\hat{C}+\hat{A}), \quad
    \phi_{\bm M_3,\sigma_-}=\frac{1}{\sqrt{2}}(\hat{A}+\hat{B}). 
\end{align}
Geometry conventions are such that $\bm M_1\to \bm M_2 \to \bm M_3$ under successive $C_{3z}$ rotations. And similarly $\hat{A}\to \hat{B} \to \hat{C}$. 

Armed with this, we can establish how $f$ behave under $C_{3z}$. For arbitrary $\bm k$ we have
\begin{align}
   C_{3z}:\,  f_{\bm k\sigma}\to f'_{\bm C_{3z}\cdot \bm k\sigma} = \sum_j\left[U_{C_{3z}}\cdot \phi_{\bm k \sigma}\right]_j c_{C_{3z}\cdot\bm k,j,\sigma}.
\end{align}
Restricting to $M$-points, this becomes simple, since $C_{3z}\cdot \bm M_\mu = \bm M_{\mu+1}$ and $\left[U_{C_{3z}}\cdot \phi_{\bm M_\mu \sigma}\right]_j=\left[\phi_{\bm M_{\mu+1} \sigma}\right]_{j}$, then
\begin{align}
   C_{3z}:\,  f_{\bm M_\mu \sigma}\to f'_{\bm M_{\mu+1}\sigma} = \sum_j\phi_{\bm M_{\mu+1} \sigma}{}_j c_{\bm M_{\mu+1},j,\sigma} = f_{\bm M_{\mu+1}\sigma}.
\end{align}
This shows that the $f$ transform trivially.

Finally, we here confirm that there are not any hidden non-trivial phase in  Eqs. \eqref{phi_ptype} and \equref{phi_mtype}. For this, we confirm that 
\begin{align}
    H(C_{3z}\cdot \bm M_\mu) = U_{C_{3z}}H(\bm M_\mu)U^\dag_{C_{3z}}
\end{align}
with 
\begin{align}
    U_{C_{3z}}&=\begin{pmatrix} 0 & 0 & 1 \\
 1 & 0 & 0 \\
 0 & 1 & 0 \end{pmatrix}.
\end{align}
Since this does not contain complex phases as matrix elements, it  confirms that there are no non-trivial phases entering \equref{phi_ptype} and \equref{phi_mtype}.

\section{Free Energy Expansion Details}

\subsection{General}
The expressions for ${\cal F}_0$, i.e. the free energy in absence of external fields, are lengthy. Instead, we provide the algorithm for generating it. First,
\begin{align}
\label{A:F0}
{\cal F}_0 = \sum_{n=1}^\infty \frac{(-1)^n}{n}\text{Tr}\left[(\hat{G} \hat{M})^n\right],
\end{align}
which follows from the expansion ${\cal F}_0 = -\text{Tr} \log (1 + \hat{G} \hat{M})$. For the case at hand, we have
\begin{align}
\left(\hat{G}\right)_{\bm k,i\omega_n}&= i\omega_n - \varepsilon_{\bm k}^c\sigma_c s_0 - \varepsilon_{\bm k}^v\sigma_v s_0, \quad \left(\hat{M}\right)_{\bm k,i\omega_n}= \bm B\cdot \bm s \sigma_0 +  g_{0,\bm k} \sigma_y+ \bm g_{\bm k}\cdot \bm s \sigma_x+ \bm d_{\bm k}\cdot \bm s \sigma_y,
\end{align}
where $\sigma_c\equiv(\sigma_0+\sigma_z)/2, \sigma_v\equiv(\sigma_0-\sigma_z)/2$.

Next, the trace acts over spin, momentum and energy. Energy trace is usual Matsubara summation, spin trace is straightforward, momentum trace requires more care so we step through it here. To account for the structure of the point group IRs, we take
\begin{align}
 g_{0,\bm k} &= \mathtt{g}_0 \cos\theta_0,  g_{x,\bm k} = \mathtt{g}_1 \sin\theta_x,  g_{y,\bm k} =- \mathtt{g}_1 \cos\theta_x,  g_{z,\bm k} = \mathtt{g}_3 \cos\theta_z,\\
  d_{x,\bm k} &= (\mathtt{d}_{1x} \cos\theta_x + \mathtt{d}_{1z} \cos\theta_z ), d_{y,\bm k} = (\mathtt{d}_{2y} \sin\theta_x + \mathtt{d}_{2z} \cos\theta_z ), d_{z,\bm k} = (\mathtt{d}_{3x} \cos\theta_x + \mathtt{d}_{3y} \sin\theta_x),
\end{align}
and implement the momentum-trace via
\begin{align}
\text{Tr}\to \int \frac{kdk}{(2\pi)}\frac{d\theta_0}{2\pi}\frac{d\theta_x}{2\pi}\frac{d\theta_z}{2\pi}.
\end{align}
Armed with this, one can readily compute ${\cal F}_0$. For numerical results shown in the main text figures, we expand to eighth order, i.e. truncate the summation in Eq.~\eqref{A:F0} at $n=n_\text{max}=8$.

\subsection{Zeeman Terms}
Denoting the Greens function for conduction and valence bands as $G^c_{\bm k,i\omega_n}$ and $G^v_{\bm k,i\omega_n}$, and with $2\hat{G}=G^c_{\bm k,i\omega_n}s_0 (\sigma_0+\sigma_z)+G^v_{\bm k,i\omega_n}s_0 (\sigma_0-\sigma_z)$, then the cubic terms work out to be
\begin{align}
    \notag {\cal F}_B^{(1)} &= \Tr [\hat{G}_{\bm k}^3(\bm d_{\bm k}\cdot \bm s)(\bm g_{\bm k}\cdot \bm s) (\bm B\cdot \bm s)] = \sum_\Gamma \sum_{m=1}^{d_\Gamma} B_i \mathtt{d}_{j,(\Gamma,m)} \mathtt{g}_{\ell} \varepsilon_{ij\ell}  T\sum_n \sum_{\bm k}  \left[\Xi_{\Gamma,m}(\bm k)\chi_{\ell}(\bm k)\left\{\left(G^c_{i\omega_n,\bm k}\right)^2  G^v_{i\omega_n,\bm k} - G^c_{i\omega_n,\bm k} \left(G^v_{i\omega_n,\bm k}\right)^2\right\}\right]\\
    \notag {\cal F}_B^{(2)} &= \Tr [\hat{G}_{\bm k}^3(\bm d_{\bm k}\cdot \bm s)(g_{0,\bm k}) (\bm B\cdot \bm s)] \hspace{0.23cm} = \sum_\Gamma \sum_{m=1}^{d_\Gamma} B_i \mathtt{d}_{j,(\Gamma,m)} \mathtt{g}_{0}\delta_{ij}T\sum_n \sum_{\bm k} \left[\Xi_{\Gamma,m}(\bm k)\chi_{0}(\bm k)\left\{\left(G^c_{i\omega_n,\bm k}\right)^2  G^v_{i\omega_n,\bm k} + G^c_{i\omega_n,\bm k} \left(G^v_{i\omega_n,\bm k}\right)^2\right\}\right]\\
    \notag {\cal F}_B^{(3)} &= \Tr [\hat{G}_{\bm k}^3(\bm d_{\bm k}\cdot \bm s)(g_{0,\bm k}) (\bm B\cdot \bm s)] \hspace{0.23cm} = \sum_\Gamma \sum_{m=1}^{d_\Gamma}B_i \mathtt{d}_{0,(\Gamma,m)} \mathtt{g}_{i}\delta_{ij}T\sum_n \sum_{\bm k} \left[\Xi_{\Gamma,m}(\bm k)\chi_{i}(\bm k)\left\{\left(G^c_{i\omega_n,\bm k}\right)^2  G^v_{i\omega_n,\bm k} + G^c_{i\omega_n,\bm k} \left(G^v_{i\omega_n,\bm k}\right)^2\right\}\right].
\end{align}
We note that in the case of spectral particle-hole symmetry, i.e. that $\varepsilon^c_{\bm k}=-\varepsilon^v_{\bm k}$, only ${\cal F}_B^{(1)}$ is nonzero.

\section{Disproving alternate scenarios}

\begin{figure}
    \centering
    \includegraphics[width=0.5\linewidth]{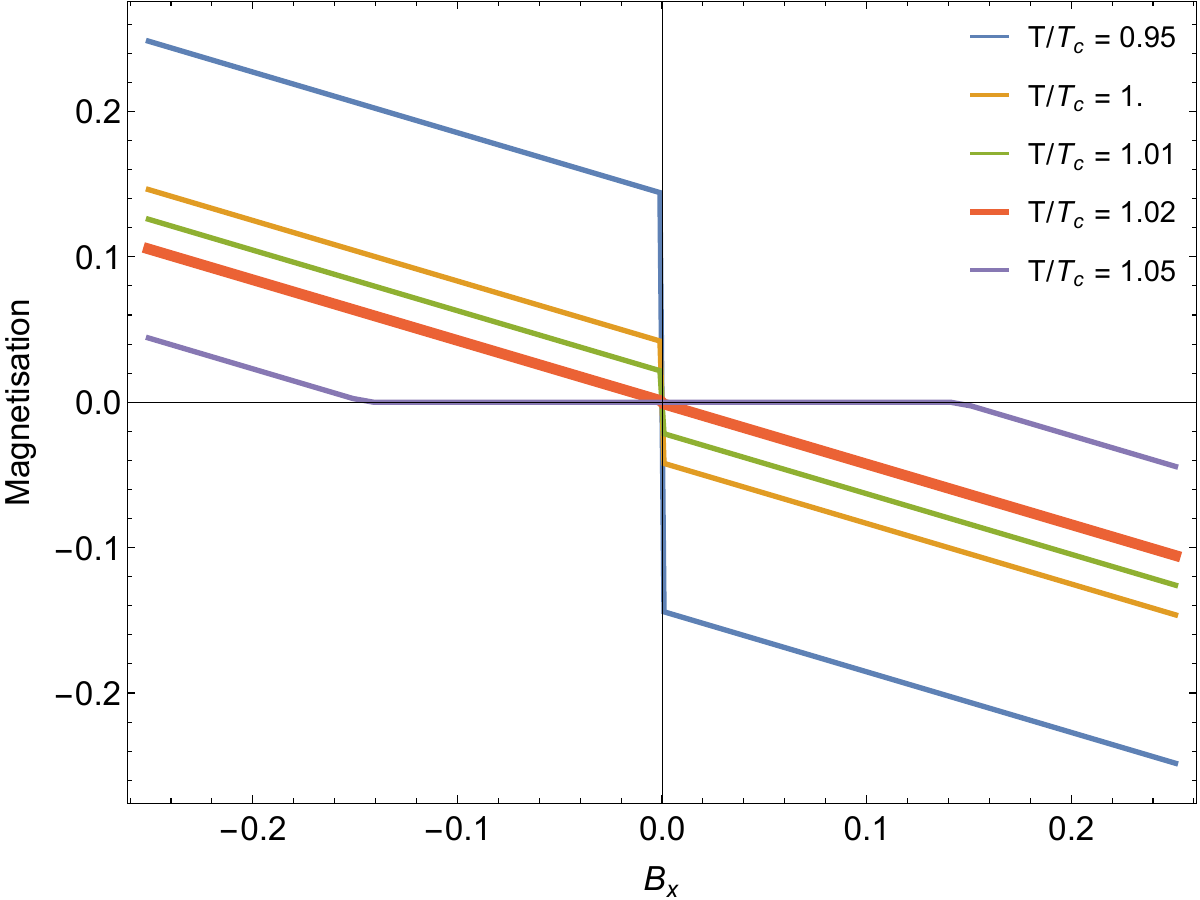}
    \caption{magnetisation from CDW. Only for a fine tuned temperature $T/T_c\approx 1.02$ (for the given choice of parameters) do we get a linear response. }
    \label{fig:CDW_scenario}
\end{figure}

\subsection{Fluctuating CDWs}
We consider the possibility that nearly critical CDW fluctuations can explain the torque. To simplify, we work with respect to point group $C_{6v}$, in that case the commensurate CDWs are classified as the IRs $F_i$ with $i=1,2,3,4$ \cite{WagnerPRB2023}. Just briefly: $F_1$ is the trivial CDW, while $F_{2,3,4}$ are odd under either/both vertical and horizontal mirror reflections. It is understood that the CDW setting at $T_\text{CDW}$ belongs to $F_1$. 

For demonstration, we consider a particular combination of $F_i$ which allows for coupling to in-plane $\bm B=(B_x,B_y,0)$,
\begin{align}
    {\cal F}_B&=g B_x \bm \Phi_{F_2} \Lambda_1 \bm \Phi_{F_3} + g B_y \bm \Phi_{F_2} \Lambda_2 \bm \Phi_{F_3}
\end{align}
here $g$ is the coupling constant, the e.g.  $\bm \Phi_{F_2} = (\Phi^{M_1}_{F_2},\Phi^{M_2}_{F_2},\Phi^{M_3}_{F_2})$ are three-component vectors, one component for each $M$-point ordering direction, and $\Lambda_1=$Diagonal$(2,-1,-1)/\sqrt{6}$, and $\Lambda_2=$Diagonal$(0,1,-1)/\sqrt{2}$.  These $\Lambda_\mu$ allow for two-dimensional, translationally invariant IRs to be formed out of the three-dimensional, CDW IRs $F_i$. 

Moreover, to explain the torque measurements, it proves crucial to include strain. In this case, $\Phi_{F_1},\bm \Phi_{F_3}$ couple to strain via
\begin{align}
    {\cal F}_\varepsilon&=\gamma_2 \varepsilon_\mu \bm \Phi_{F_2} \Lambda_\mu \bm \Phi_{F_2}+\gamma_3 \varepsilon_\mu \bm \Phi_{F_3} \Lambda_\mu \bm \Phi_{F_3}.
\end{align}
Here $\gamma_{2,3}$ are coupling constants. 
We again make use of the $\Lambda_\mu$. But note, now e.g. $\bm \Phi_{F_2} \Lambda_\mu \bm \Phi_{F_2} \in E_2$ whereas $\bm \Phi_{F_2} \Lambda_\mu \bm \Phi_{F_3} \in E_1$, due to the difference in mirror symmetries of $\bm \Phi_{F_2}$ and $\bm \Phi_{F_3}$. 

Finally, we treat the zero-field and strain free energy as
\begin{align}
    {\cal F}_0&=s_2 \bm \Phi_{F_2}.\bm \Phi_{F_2} + s_3 \bm \Phi_{F_3}.\bm \Phi_{F_3} + \lambda_2 (\bm \Phi_{F_2}.\bm \Phi_{F_2})^2 + \lambda_3 (\bm \Phi_{F_3}.\bm \Phi_{F_3})^2.
\end{align}
The $s_{2,3}$ tune the phase transition, for simplicity in our modeling we take
\begin{align}
    s_2=s_3\sim(T-T_c),
\end{align}
and set $\lambda_2=\lambda_3$. 

\textbf{The upshot---} This detailed modelling serves to illustrate what may already be intuitively clear: since the magnetic field couples only \textit{quadratically} (not linearly) to the CDW order parameters, a linear-in-$B$ response requires fine tuning---specifically, setting $s_2 = s_3 = 0$. Minimisation of the free energy ${\cal F} = {\cal F}_0 + {\cal F}_B + {\cal F}_\epsilon$, using representative parameters, confirms this expectation: linear response emerges only under fine-tuned conditions (see Fig.~\ref{fig:CDW_scenario}). However, the experimental data (Fig.~2a of Ref. \cite{Asaba2024}) shows a robust linear response across a broad temperature range, $60 \leq T \lesssim 130\,\text{K}$, which includes the onset of the CDW phase. We therefore conclude that the CDW scenario is highly unlikely.

\subsection{Single band magnetisation}\label{IntrabandTerms}
\subsubsection{Spin magnetism}
In a single-band model with spinful electrons, a magnetic field can couple linearly to a spin-1 order parameter, giving rise to a linear-in-$B$ response,
\begin{align}
    {\cal F}_B &= g\, \bm{B} \cdot \bm{\Phi}
\end{align}
where \(\bm{\Phi} = \langle \sum_{\bm{k}} c^\dagger_{\bm{k}} \bm{\sigma} c_{\bm{k}} \rangle\) represents a uniform spin magnetisation. However, in the absence of spin-orbit coupling (SOC), the spin susceptibility remains isotropic, and there is no mechanism for strain to couple to spin in a way that induces magnetic anisotropy. Consequently, no torque response can arise. For the point group $D_{6h}$, SOC within a single band is forbidden by symmetry, so such anisotropic responses require either multi-band effects or explicit symmetry breaking.

\begin{figure}[t]
    \centering
    \includegraphics[width=0.5\linewidth]{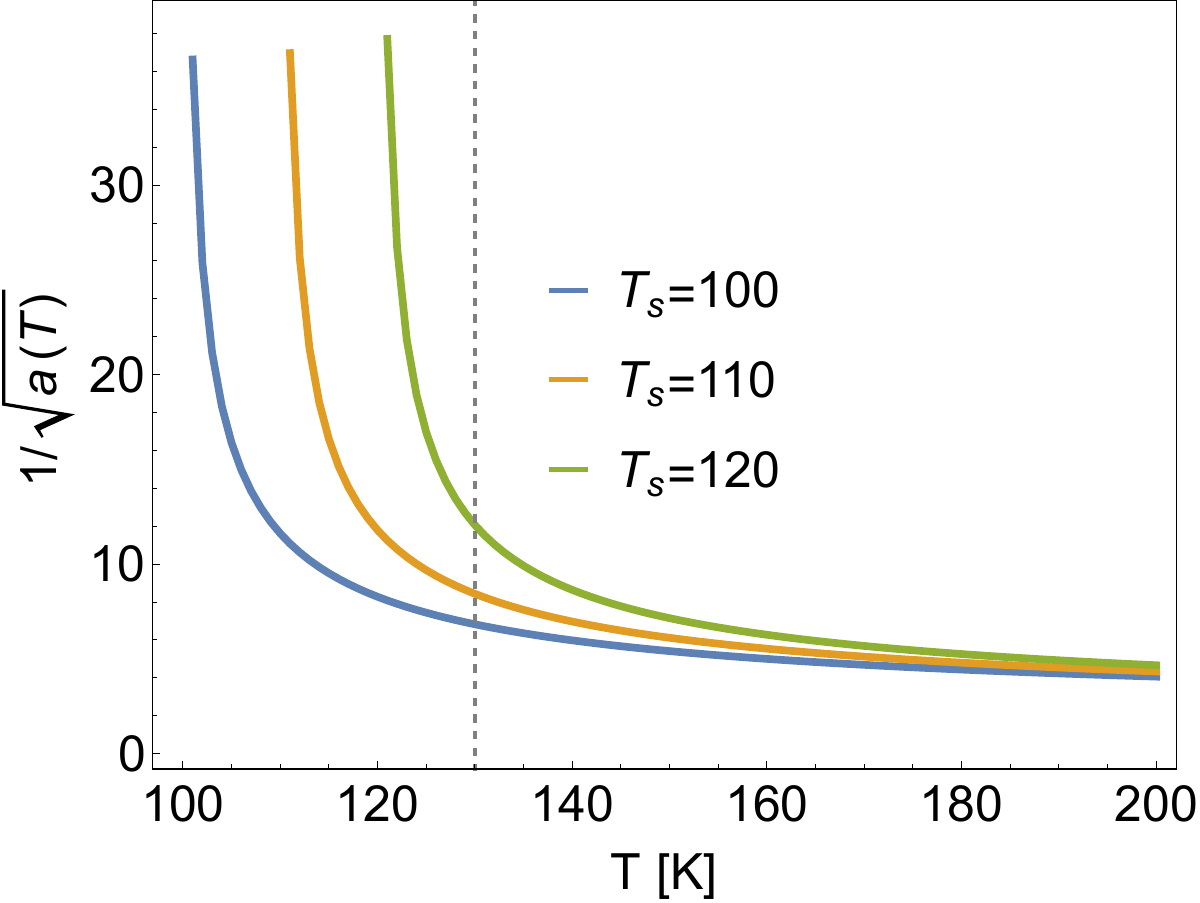}
    \caption{Stoner-like enhancement}
    \label{fig:stoner}
\end{figure}

\subsubsection{Orbital magnetism}
Intraband orbital magnetism may indeed couple to strain. However, we detail of few reasons why this scenario is unlikely:
\begin{itemize}
    \item {\bf Vanishes in 2D limit.} In-plane orbital magnetism implies loops of current along the $z$-axis. For kagome metals AV$_3$Sb$_5$, $z$-axis hopping is significantly suppressed relative to in-plane hopping; this follows from the layered structure of AV$_3$Sb$_5$, together with the large transport anisotropy, specifically for the case of interest CsV$_3$Sb$_5$, of $\rho_c/\rho_{ab}=600$ \cite{OrtizTeicherPRL2020}. As a result, we expect a vanishingly small orbital $g$-factor and hence a vanishingly small paramagnetic response.  

     \item {\bf Crossover.} One could envisage that the orbital magnetic susceptibility $\chi_{\text{orb}}(T)$ is thermally renormalised, and that the $g$-factor is correspondingly enhanced via
\begin{align}
    g(T) \propto \sqrt{\chi_{\text{orb}}(T)}.
\end{align}
To describe this from a free energy perspective, we consider the Landau expansion in the orbital magnetic order parameter $\mathbf{P}$:
\begin{align}
    \mathcal{F}[\mathbf{P}] = \frac{a(T)}{2} \, \mathbf{P}^2 + \frac{b}{4} (\mathbf{P}^2)^2 + \cdots - \mathbf{B} \cdot \mathbf{P},
\end{align}
with $a(T) \propto 1/\chi_{\text{orb}}(T)$. As temperature decreases, interactions can enhance $\chi_{\text{orb}}(T)$, driving $a(T) \to 0$, which signals an instability to spontaneous orbital magnetism (a Stoner-like transition). 

To match experimental observations, where $g(T)$ increases sharply at a characteristic temperature $T_\tau$ but the system avoids developing long-range order, one would need a highly fine-tuned behaviour of $a(T)$: it must remain nearly constant at high $T$, then sharply decrease near $T_\tau$ (causing $g(T)$ to rise), but somehow saturate without reaching $a(T)=0$ — thus narrowly avoiding the instability. We view this as an implausible tuning.

To make this more concrete, we compute $a(T)$ within mean-field theory, which is given by
\begin{align}
    a(T) = \frac{1}{V_\text{eff}} - \int \frac{d^2k}{(2\pi)^2} \frac{\text{sech}^2\left( \frac{\varepsilon_{\bm k}}{2T} \right)}{4T},
\end{align}
where $\varepsilon_{\bm k}$ is the conduction band dispersion and the choice of $V_\text{eff}$ sets the Stoner transition temperature $T_s$.  Evaluating this expression shows that $1/\sqrt{a(T)}$ varies smoothly with temperature and exhibits no abrupt enhancement except in the vicinity of a Stoner instability, where $a(T_s)=0$. Plots for representative values of $T_s$ are shown in Fig.~\ref{fig:stoner}.

\end{itemize}

\end{widetext}
\end{appendix}

\begin{thebibliography}{42}%
\makeatletter
\providecommand \@ifxundefined [1]{%
 \@ifx{#1\undefined}
}%
\providecommand \@ifnum [1]{%
 \ifnum #1\expandafter \@firstoftwo
 \else \expandafter \@secondoftwo
 \fi
}%
\providecommand \@ifx [1]{%
 \ifx #1\expandafter \@firstoftwo
 \else \expandafter \@secondoftwo
 \fi
}%
\providecommand \natexlab [1]{#1}%
\providecommand \enquote  [1]{``#1''}%
\providecommand \bibnamefont  [1]{#1}%
\providecommand \bibfnamefont [1]{#1}%
\providecommand \citenamefont [1]{#1}%
\providecommand \href@noop [0]{\@secondoftwo}%
\providecommand \href [0]{\begingroup \@sanitize@url \@href}%
\providecommand \@href[1]{\@@startlink{#1}\@@href}%
\providecommand \@@href[1]{\endgroup#1\@@endlink}%
\providecommand \@sanitize@url [0]{\catcode `\\12\catcode `\$12\catcode
  `\&12\catcode `\#12\catcode `\^12\catcode `\_12\catcode `\%12\relax}%
\providecommand \@@startlink[1]{}%
\providecommand \@@endlink[0]{}%
\providecommand \url  [0]{\begingroup\@sanitize@url \@url }%
\providecommand \@url [1]{\endgroup\@href {#1}{\urlprefix }}%
\providecommand \urlprefix  [0]{URL }%
\providecommand \Eprint [0]{\href }%
\providecommand \doibase [0]{http://dx.doi.org/}%
\providecommand \selectlanguage [0]{\@gobble}%
\providecommand \bibinfo  [0]{\@secondoftwo}%
\providecommand \bibfield  [0]{\@secondoftwo}%
\providecommand \translation [1]{[#1]}%
\providecommand \BibitemOpen [0]{}%
\providecommand \bibitemStop [0]{}%
\providecommand \bibitemNoStop [0]{.\EOS\space}%
\providecommand \EOS [0]{\spacefactor3000\relax}%
\providecommand \BibitemShut  [1]{\csname bibitem#1\endcsname}%
\let\auto@bib@innerbib\@empty
\bibitem [{\citenamefont {Wang}\ \emph {et~al.}(2023)\citenamefont {Wang},
  \citenamefont {Wu}, \citenamefont {McCandless}, \citenamefont {Chan},\ and\
  \citenamefont {Ali}}]{wang2023quantum}%
  \BibitemOpen
  \bibfield  {author} {\bibinfo {author} {\bibfnamefont {Y.}~\bibnamefont
  {Wang}}, \bibinfo {author} {\bibfnamefont {H.}~\bibnamefont {Wu}}, \bibinfo
  {author} {\bibfnamefont {G.~T.}\ \bibnamefont {McCandless}}, \bibinfo
  {author} {\bibfnamefont {J.~Y.}\ \bibnamefont {Chan}}, \ and\ \bibinfo
  {author} {\bibfnamefont {M.~N.}\ \bibnamefont {Ali}},\ }\bibfield  {title}
  {\enquote {\bibinfo {title} {Quantum states and intertwining phases in kagome
  materials},}\ }\href {https://www.nature.com/articles/s42254-023-00635-7}
  {\bibfield  {journal} {\bibinfo  {journal} {Nature Reviews Physics}\ }\textbf
  {\bibinfo {volume} {5}},\ \bibinfo {pages} {635} (\bibinfo {year}
  {2023})}\BibitemShut {NoStop}%
\bibitem [{\citenamefont {Yin}\ \emph {et~al.}(2022)\citenamefont {Yin},
  \citenamefont {Lian},\ and\ \citenamefont {Hasan}}]{yin2022topological}%
  \BibitemOpen
  \bibfield  {author} {\bibinfo {author} {\bibfnamefont {J.-X.}\ \bibnamefont
  {Yin}}, \bibinfo {author} {\bibfnamefont {B.}~\bibnamefont {Lian}}, \ and\
  \bibinfo {author} {\bibfnamefont {M.~Z.}\ \bibnamefont {Hasan}},\ }\bibfield
  {title} {\enquote {\bibinfo {title} {Topological kagome magnets and
  superconductors},}\ }\href
  {https://www.nature.com/articles/s41586-022-05516-0} {\bibfield  {journal}
  {\bibinfo  {journal} {Nature}\ }\textbf {\bibinfo {volume} {612}},\ \bibinfo
  {pages} {647} (\bibinfo {year} {2022})}\BibitemShut {NoStop}%
\bibitem [{\citenamefont {Neupert}\ \emph {et~al.}(2022)\citenamefont
  {Neupert}, \citenamefont {Denner}, \citenamefont {Yin}, \citenamefont
  {Thomale},\ and\ \citenamefont {Hasan}}]{neupert2022review}%
  \BibitemOpen
  \bibfield  {author} {\bibinfo {author} {\bibfnamefont {T.}~\bibnamefont
  {Neupert}}, \bibinfo {author} {\bibfnamefont {M.~M.}\ \bibnamefont {Denner}},
  \bibinfo {author} {\bibfnamefont {J.-X.}\ \bibnamefont {Yin}}, \bibinfo
  {author} {\bibfnamefont {R.}~\bibnamefont {Thomale}}, \ and\ \bibinfo
  {author} {\bibfnamefont {M.~Z.}\ \bibnamefont {Hasan}},\ }\bibfield  {title}
  {\enquote {\bibinfo {title} {Charge order and superconductivity in kagome
  materials},}\ }\href {https://www.nature.com/articles/s41567-021-01404-y}
  {\bibfield  {journal} {\bibinfo  {journal} {Nat. Phys.}\ }\textbf {\bibinfo
  {volume} {18}},\ \bibinfo {pages} {137} (\bibinfo {year} {2022})}\BibitemShut
  {NoStop}%
\bibitem [{\citenamefont {Ortiz}\ \emph {et~al.}(2019)\citenamefont {Ortiz},
  \citenamefont {Gomes}, \citenamefont {Morey}, \citenamefont {Winiarski},
  \citenamefont {Bordelon}, \citenamefont {Mangum}, \citenamefont {Oswald},
  \citenamefont {Rodriguez-Rivera}, \citenamefont {Neilson}, \citenamefont
  {Wilson}, \citenamefont {Ertekin}, \citenamefont {McQueen},\ and\
  \citenamefont {Toberer}}]{PhysRevMaterials.3.094407}%
  \BibitemOpen
  \bibfield  {author} {\bibinfo {author} {\bibfnamefont {B.~R.}\ \bibnamefont
  {Ortiz}}, \bibinfo {author} {\bibfnamefont {L.~C.}\ \bibnamefont {Gomes}},
  \bibinfo {author} {\bibfnamefont {J.~R.}\ \bibnamefont {Morey}}, \bibinfo
  {author} {\bibfnamefont {M.}~\bibnamefont {Winiarski}}, \bibinfo {author}
  {\bibfnamefont {M.}~\bibnamefont {Bordelon}}, \bibinfo {author}
  {\bibfnamefont {J.~S.}\ \bibnamefont {Mangum}}, \bibinfo {author}
  {\bibfnamefont {I.~W.~H.}\ \bibnamefont {Oswald}}, \bibinfo {author}
  {\bibfnamefont {J.~A.}\ \bibnamefont {Rodriguez-Rivera}}, \bibinfo {author}
  {\bibfnamefont {J.~R.}\ \bibnamefont {Neilson}}, \bibinfo {author}
  {\bibfnamefont {S.~D.}\ \bibnamefont {Wilson}}, \bibinfo {author}
  {\bibfnamefont {E.}~\bibnamefont {Ertekin}}, \bibinfo {author} {\bibfnamefont
  {T.~M.}\ \bibnamefont {McQueen}}, \ and\ \bibinfo {author} {\bibfnamefont
  {E.~S.}\ \bibnamefont {Toberer}},\ }\bibfield  {title} {\enquote {\bibinfo
  {title} {New kagome prototype materials: discovery of
  {KV}$_{3}${Sb}$_{5}$,{RbV}$_{3}${Sb}$_{5}$, and {CsV}$_{3}${Sb}$_{5}$},}\
  }\href {\doibase 10.1103/PhysRevMaterials.3.094407} {\bibfield  {journal}
  {\bibinfo  {journal} {Phys. Rev. Mater.}\ }\textbf {\bibinfo {volume} {3}},\
  \bibinfo {pages} {094407} (\bibinfo {year} {2019})}\BibitemShut {NoStop}%
\bibitem [{\citenamefont {Ortiz}\ \emph
  {et~al.}(2020{\natexlab{a}})\citenamefont {Ortiz}, \citenamefont {Teicher},
  \citenamefont {Hu}, \citenamefont {Zuo}, \citenamefont {Sarte}, \citenamefont
  {Schueller}, \citenamefont {Abeykoon}, \citenamefont {Krogstad},
  \citenamefont {Rosenkranz}, \citenamefont {Osborn}, \citenamefont {Seshadri},
  \citenamefont {Balents}, \citenamefont {He},\ and\ \citenamefont
  {Wilson}}]{PhysRevLett.125.247002}%
  \BibitemOpen
  \bibfield  {author} {\bibinfo {author} {\bibfnamefont {B.~R.}\ \bibnamefont
  {Ortiz}}, \bibinfo {author} {\bibfnamefont {S.~M.~L.}\ \bibnamefont
  {Teicher}}, \bibinfo {author} {\bibfnamefont {Y.}~\bibnamefont {Hu}},
  \bibinfo {author} {\bibfnamefont {J.~L.}\ \bibnamefont {Zuo}}, \bibinfo
  {author} {\bibfnamefont {P.~M.}\ \bibnamefont {Sarte}}, \bibinfo {author}
  {\bibfnamefont {E.~C.}\ \bibnamefont {Schueller}}, \bibinfo {author}
  {\bibfnamefont {A.~M.~M.}\ \bibnamefont {Abeykoon}}, \bibinfo {author}
  {\bibfnamefont {M.~J.}\ \bibnamefont {Krogstad}}, \bibinfo {author}
  {\bibfnamefont {S.}~\bibnamefont {Rosenkranz}}, \bibinfo {author}
  {\bibfnamefont {R.}~\bibnamefont {Osborn}}, \bibinfo {author} {\bibfnamefont
  {R.}~\bibnamefont {Seshadri}}, \bibinfo {author} {\bibfnamefont
  {L.}~\bibnamefont {Balents}}, \bibinfo {author} {\bibfnamefont
  {J.}~\bibnamefont {He}}, \ and\ \bibinfo {author} {\bibfnamefont {S.~D.}\
  \bibnamefont {Wilson}},\ }\bibfield  {title} {\enquote {\bibinfo {title}
  {{CsV}$_3${Sb}$_5$: A $\mathbb{Z}_{2}$ topological kagome metal with a
  superconducting ground state},}\ }\href
  {https://link.aps.org/doi/10.1103/PhysRevLett.125.247002} {\bibfield
  {journal} {\bibinfo  {journal} {Phys. Rev. Lett.}\ }\textbf {\bibinfo
  {volume} {125}},\ \bibinfo {pages} {247002} (\bibinfo {year}
  {2020}{\natexlab{a}})}\BibitemShut {NoStop}%
\bibitem [{\citenamefont {Yang}\ \emph {et~al.}(2020)\citenamefont {Yang} \emph
  {et~al.}}]{yang2021giant}%
  \BibitemOpen
  \bibfield  {author} {\bibinfo {author} {\bibfnamefont {S.-Y.}\ \bibnamefont
  {Yang}} \emph {et~al.},\ }\bibfield  {title} {\enquote {\bibinfo {title}
  {Giant, unconventional anomalous {Hall} effect in the metallic frustrated
  magnet candidate, {KV}$_3${Sb}$_5$},}\ }\href
  {https://www.science.org/doi/10.1126/sciadv.abb6003} {\bibfield  {journal}
  {\bibinfo  {journal} {Science Advances}\ }\textbf {\bibinfo {volume} {6}},\
  \bibinfo {pages} {eabb6003} (\bibinfo {year} {2020})}\BibitemShut {NoStop}%
\bibitem [{\citenamefont {Chen}\ \emph {et~al.}(2022)\citenamefont {Chen} \emph
  {et~al.}}]{Chen2022anomalous}%
  \BibitemOpen
  \bibfield  {author} {\bibinfo {author} {\bibfnamefont {D.}~\bibnamefont
  {Chen}} \emph {et~al.},\ }\bibfield  {title} {\enquote {\bibinfo {title}
  {Anomalous thermoelectric effects and quantum oscillations in the kagome
  metal {CsV}$_3${Sb}$_5$},}\ }\href
  {https://journals.aps.org/prb/abstract/10.1103/PhysRevB.105.L201109}
  {\bibfield  {journal} {\bibinfo  {journal} {Phys. Rev. B}\ }\textbf {\bibinfo
  {volume} {105}},\ \bibinfo {pages} {L201109} (\bibinfo {year}
  {2022})}\BibitemShut {NoStop}%
\bibitem [{\citenamefont {Mielke}\ \emph {et~al.}(2022)\citenamefont {Mielke}
  \emph {et~al.}}]{Mielke2021b}%
  \BibitemOpen
  \bibfield  {author} {\bibinfo {author} {\bibfnamefont {C.}~\bibnamefont
  {Mielke}} \emph {et~al.},\ }\bibfield  {title} {\enquote {\bibinfo {title}
  {Time-reversal symmetry-breaking charge order in a kagome superconductor},}\
  }\href {https://www.nature.com/articles/s41586-021-04327-z} {\bibfield
  {journal} {\bibinfo  {journal} {Nature}\ }\textbf {\bibinfo {volume} {602}},\
  \bibinfo {pages} {245–250} (\bibinfo {year} {2022})}\BibitemShut {NoStop}%
\bibitem [{\citenamefont {Jiang}\ \emph {et~al.}(2021)\citenamefont {Jiang},
  \citenamefont {Yin}, \citenamefont {Denner}, \citenamefont {Shumiya},
  \citenamefont {Ortiz}, \citenamefont {Xu}, \citenamefont {Guguchia},
  \citenamefont {He}, \citenamefont {Hossain}, \citenamefont {Liu},
  \citenamefont {Ruff}, \citenamefont {Kautzsch}, \citenamefont {Zhang},
  \citenamefont {Chang}, \citenamefont {Belopolski}, \citenamefont {Zhang},
  \citenamefont {Cochran}, \citenamefont {Multer}, \citenamefont {Litskevich},
  \citenamefont {Cheng}, \citenamefont {Yang}, \citenamefont {Wang},
  \citenamefont {Thomale}, \citenamefont {Neupert}, \citenamefont {Wilson},\
  and\ \citenamefont {Hasan}}]{jiang2021np}%
  \BibitemOpen
  \bibfield  {author} {\bibinfo {author} {\bibfnamefont {Y.-X.}\ \bibnamefont
  {Jiang}}, \bibinfo {author} {\bibfnamefont {J.-X.}\ \bibnamefont {Yin}},
  \bibinfo {author} {\bibfnamefont {M.~M.}\ \bibnamefont {Denner}}, \bibinfo
  {author} {\bibfnamefont {N.}~\bibnamefont {Shumiya}}, \bibinfo {author}
  {\bibfnamefont {B.~R.}\ \bibnamefont {Ortiz}}, \bibinfo {author}
  {\bibfnamefont {G.}~\bibnamefont {Xu}}, \bibinfo {author} {\bibfnamefont
  {Z.}~\bibnamefont {Guguchia}}, \bibinfo {author} {\bibfnamefont
  {J.}~\bibnamefont {He}}, \bibinfo {author} {\bibfnamefont {M.~S.}\
  \bibnamefont {Hossain}}, \bibinfo {author} {\bibfnamefont {X.}~\bibnamefont
  {Liu}}, \bibinfo {author} {\bibfnamefont {J.}~\bibnamefont {Ruff}}, \bibinfo
  {author} {\bibfnamefont {L.}~\bibnamefont {Kautzsch}}, \bibinfo {author}
  {\bibfnamefont {S.~S.}\ \bibnamefont {Zhang}}, \bibinfo {author}
  {\bibfnamefont {G.}~\bibnamefont {Chang}}, \bibinfo {author} {\bibfnamefont
  {I.}~\bibnamefont {Belopolski}}, \bibinfo {author} {\bibfnamefont
  {Q.}~\bibnamefont {Zhang}}, \bibinfo {author} {\bibfnamefont {T.~A.}\
  \bibnamefont {Cochran}}, \bibinfo {author} {\bibfnamefont {D.}~\bibnamefont
  {Multer}}, \bibinfo {author} {\bibfnamefont {M.}~\bibnamefont {Litskevich}},
  \bibinfo {author} {\bibfnamefont {Z.-J.}\ \bibnamefont {Cheng}}, \bibinfo
  {author} {\bibfnamefont {X.~P.}\ \bibnamefont {Yang}}, \bibinfo {author}
  {\bibfnamefont {Z.}~\bibnamefont {Wang}}, \bibinfo {author} {\bibfnamefont
  {R.}~\bibnamefont {Thomale}}, \bibinfo {author} {\bibfnamefont
  {T.}~\bibnamefont {Neupert}}, \bibinfo {author} {\bibfnamefont {S.~D.}\
  \bibnamefont {Wilson}}, \ and\ \bibinfo {author} {\bibfnamefont {M.~Z.}\
  \bibnamefont {Hasan}},\ }\bibfield  {title} {\enquote {\bibinfo {title}
  {Unconventional chiral charge order in kagome superconductor
  {KV}$_3${Sb}$_5$},}\ }\href
  {https://www.nature.com/articles/s41563-021-01034-y} {\bibfield  {journal}
  {\bibinfo  {journal} {Nat. Mater.}\ }\textbf {\bibinfo {volume} {20}},\
  \bibinfo {pages} {1353} (\bibinfo {year} {2021})}\BibitemShut {NoStop}%
\bibitem [{\citenamefont {Xu}\ \emph {et~al.}(2022)\citenamefont {Xu},
  \citenamefont {Ni}, \citenamefont {Liu}, \citenamefont {Ortiz}, \citenamefont
  {Deng}, \citenamefont {Wilson}, \citenamefont {Yan}, \citenamefont
  {Balents},\ and\ \citenamefont {Wu}}]{Xu2022}%
  \BibitemOpen
  \bibfield  {author} {\bibinfo {author} {\bibfnamefont {Y.}~\bibnamefont
  {Xu}}, \bibinfo {author} {\bibfnamefont {Z.}~\bibnamefont {Ni}}, \bibinfo
  {author} {\bibfnamefont {Y.}~\bibnamefont {Liu}}, \bibinfo {author}
  {\bibfnamefont {B.~R.}\ \bibnamefont {Ortiz}}, \bibinfo {author}
  {\bibfnamefont {Q.}~\bibnamefont {Deng}}, \bibinfo {author} {\bibfnamefont
  {S.~D.}\ \bibnamefont {Wilson}}, \bibinfo {author} {\bibfnamefont
  {B.}~\bibnamefont {Yan}}, \bibinfo {author} {\bibfnamefont {L.}~\bibnamefont
  {Balents}}, \ and\ \bibinfo {author} {\bibfnamefont {L.}~\bibnamefont {Wu}},\
  }\bibfield  {title} {\enquote {\bibinfo {title} {Three-state nematicity and
  magneto-optical {Kerr} effect in the charge density waves in kagome
  superconductors},}\ }\href
  {https://www.nature.com/articles/s41567-022-01805-7} {\bibfield  {journal}
  {\bibinfo  {journal} {Nat. Phys.}\ }\textbf {\bibinfo {volume} {18}},\
  \bibinfo {pages} {1470} (\bibinfo {year} {2022})}\BibitemShut {NoStop}%
\bibitem [{\citenamefont {Li}\ \emph {et~al.}(2022)\citenamefont {Li} \emph
  {et~al.}}]{li2021rotation}%
  \BibitemOpen
  \bibfield  {author} {\bibinfo {author} {\bibfnamefont {H.}~\bibnamefont {Li}}
  \emph {et~al.},\ }\bibfield  {title} {\enquote {\bibinfo {title} {Rotation
  symmetry breaking in the normal state of a kagome superconductor
  {KV}$_3${Sb}$_5$},}\ }\href
  {https://www.nature.com/articles/s41567-021-01479-7} {\bibfield  {journal}
  {\bibinfo  {journal} {Nat. Phys.}\ }\textbf {\bibinfo {volume} {18}},\
  \bibinfo {pages} {265} (\bibinfo {year} {2022})}\BibitemShut {NoStop}%
\bibitem [{\citenamefont {Saykin}\ \emph {et~al.}(2023)\citenamefont {Saykin}
  \emph {et~al.}}]{saykin2023high}%
  \BibitemOpen
  \bibfield  {author} {\bibinfo {author} {\bibfnamefont {D.~R.}\ \bibnamefont
  {Saykin}} \emph {et~al.},\ }\bibfield  {title} {\enquote {\bibinfo {title}
  {High resolution polar {Kerr} effect studies of {CsV}$_{3}${Sb}$_{5}$: Tests
  for time-reversal symmetry breaking below the charge-order transition},}\
  }\href {https://journals.aps.org/prl/abstract/10.1103/PhysRevLett.131.016901}
  {\bibfield  {journal} {\bibinfo  {journal} {Phys. Rev. Lett.}\ }\textbf
  {\bibinfo {volume} {131}},\ \bibinfo {pages} {016901} (\bibinfo {year}
  {2023})}\BibitemShut {NoStop}%
\bibitem [{\citenamefont {Zhao}\ \emph {et~al.}(2021)\citenamefont {Zhao},
  \citenamefont {Li}, \citenamefont {Ortiz}, \citenamefont {Teicher},
  \citenamefont {Park}, \citenamefont {Ye}, \citenamefont {Wang}, \citenamefont
  {Balents}, \citenamefont {Wilson},\ and\ \citenamefont
  {Zeljkovic}}]{zhao2021cascade}%
  \BibitemOpen
  \bibfield  {author} {\bibinfo {author} {\bibfnamefont {H.}~\bibnamefont
  {Zhao}}, \bibinfo {author} {\bibfnamefont {H.}~\bibnamefont {Li}}, \bibinfo
  {author} {\bibfnamefont {B.~R.}\ \bibnamefont {Ortiz}}, \bibinfo {author}
  {\bibfnamefont {S.~M.}\ \bibnamefont {Teicher}}, \bibinfo {author}
  {\bibfnamefont {T.}~\bibnamefont {Park}}, \bibinfo {author} {\bibfnamefont
  {M.}~\bibnamefont {Ye}}, \bibinfo {author} {\bibfnamefont {Z.}~\bibnamefont
  {Wang}}, \bibinfo {author} {\bibfnamefont {L.}~\bibnamefont {Balents}},
  \bibinfo {author} {\bibfnamefont {S.~D.}\ \bibnamefont {Wilson}}, \ and\
  \bibinfo {author} {\bibfnamefont {I.}~\bibnamefont {Zeljkovic}},\ }\bibfield
  {title} {\enquote {\bibinfo {title} {Cascade of correlated electron states in
  the kagome superconductor {CsV}$_3${Sb}$_5$},}\ }\href
  {https://www.nature.com/articles/s41586-021-03946-w} {\bibfield  {journal}
  {\bibinfo  {journal} {Nature}\ }\textbf {\bibinfo {volume} {599}},\ \bibinfo
  {pages} {216} (\bibinfo {year} {2021})}\BibitemShut {NoStop}%
\bibitem [{\citenamefont {Xu}\ \emph {et~al.}(2021)\citenamefont {Xu} \emph
  {et~al.}}]{Xu2021}%
  \BibitemOpen
  \bibfield  {author} {\bibinfo {author} {\bibfnamefont {H.-S.}\ \bibnamefont
  {Xu}} \emph {et~al.},\ }\bibfield  {title} {\enquote {\bibinfo {title}
  {Multiband superconductivity with sign-preserving order parameter in kagome
  superconductor {CsV}$_3${Sb}$_5$},}\ }\href
  {https://journals.aps.org/prl/abstract/10.1103/PhysRevLett.127.187004}
  {\bibfield  {journal} {\bibinfo  {journal} {Phys. Rev. Lett.}\ }\textbf
  {\bibinfo {volume} {127}},\ \bibinfo {pages} {187004} (\bibinfo {year}
  {2021})}\BibitemShut {NoStop}%
\bibitem [{\citenamefont {Gupta}\ \emph {et~al.}(2022)\citenamefont {Gupta}
  \emph {et~al.}}]{Gupta2021}%
  \BibitemOpen
  \bibfield  {author} {\bibinfo {author} {\bibfnamefont {R.}~\bibnamefont
  {Gupta}} \emph {et~al.},\ }\bibfield  {title} {\enquote {\bibinfo {title}
  {Microscopic evidence for anisotropic multigap superconductivity in the
  {CsV}$_3${Sb}$_5$ kagome superconductor},}\ }\href
  {https://www.nature.com/articles/s41535-022-00453-7} {\bibfield  {journal}
  {\bibinfo  {journal} {npj Quantum Materials}\ }\textbf {\bibinfo {volume}
  {7}},\ \bibinfo {pages} {1} (\bibinfo {year} {2022})}\BibitemShut {NoStop}%
\bibitem [{\citenamefont {Duan}\ \emph {et~al.}(2021)\citenamefont {Duan} \emph
  {et~al.}}]{Duan2021}%
  \BibitemOpen
  \bibfield  {author} {\bibinfo {author} {\bibfnamefont {W.}~\bibnamefont
  {Duan}} \emph {et~al.},\ }\bibfield  {title} {\enquote {\bibinfo {title}
  {Nodeless superconductivity in the kagome metal {CsV}$_3${Sb}$_5$},}\ }\href
  {https://link.springer.com/article/10.1007/s11433-021-1747-7} {\bibfield
  {journal} {\bibinfo  {journal} {Science China Physics, Mechanics \&
  Astronomy}\ }\textbf {\bibinfo {volume} {64}},\ \bibinfo {pages} {1}
  (\bibinfo {year} {2021})}\BibitemShut {NoStop}%
\bibitem [{\citenamefont {Li}\ \emph {et~al.}(2021)\citenamefont {Li},
  \citenamefont {Zhang}, \citenamefont {Yilmaz}, \citenamefont {Pai},
  \citenamefont {Marvinney}, \citenamefont {Said}, \citenamefont {Yin},
  \citenamefont {Gong}, \citenamefont {Tu}, \citenamefont {Vescovo} \emph
  {et~al.}}]{Li2021c}%
  \BibitemOpen
  \bibfield  {author} {\bibinfo {author} {\bibfnamefont {H.}~\bibnamefont
  {Li}}, \bibinfo {author} {\bibfnamefont {T.}~\bibnamefont {Zhang}}, \bibinfo
  {author} {\bibfnamefont {T.}~\bibnamefont {Yilmaz}}, \bibinfo {author}
  {\bibfnamefont {Y.}~\bibnamefont {Pai}}, \bibinfo {author} {\bibfnamefont
  {C.}~\bibnamefont {Marvinney}}, \bibinfo {author} {\bibfnamefont
  {A.}~\bibnamefont {Said}}, \bibinfo {author} {\bibfnamefont {Q.}~\bibnamefont
  {Yin}}, \bibinfo {author} {\bibfnamefont {C.}~\bibnamefont {Gong}}, \bibinfo
  {author} {\bibfnamefont {Z.}~\bibnamefont {Tu}}, \bibinfo {author}
  {\bibfnamefont {E.}~\bibnamefont {Vescovo}},  \emph {et~al.},\ }\bibfield
  {title} {\enquote {\bibinfo {title} {Observation of unconventional charge
  density wave without acoustic phonon anomaly in kagome superconductors
  {AV}$_3${Sb}$_5$ ({A}= {Rb, Cs})},}\ }\href
  {https://journals.aps.org/prx/abstract/10.1103/PhysRevX.11.031050} {\bibfield
   {journal} {\bibinfo  {journal} {Physical Review X}\ }\textbf {\bibinfo
  {volume} {11}},\ \bibinfo {pages} {031050} (\bibinfo {year}
  {2021})}\BibitemShut {NoStop}%
\bibitem [{\citenamefont {Shumiya}\ \emph {et~al.}(2021)\citenamefont {Shumiya}
  \emph {et~al.}}]{Shumiya2021}%
  \BibitemOpen
  \bibfield  {author} {\bibinfo {author} {\bibfnamefont {N.}~\bibnamefont
  {Shumiya}} \emph {et~al.},\ }\bibfield  {title} {\enquote {\bibinfo {title}
  {Intrinsic nature of chiral charge order in the kagome superconductor
  {RbV}$_3${Sb}$_5$},}\ }\href
  {https://journals.aps.org/prb/abstract/10.1103/PhysRevB.104.035131}
  {\bibfield  {journal} {\bibinfo  {journal} {Phys. Rev. B}\ }\textbf {\bibinfo
  {volume} {104}},\ \bibinfo {pages} {035131} (\bibinfo {year}
  {2021})}\BibitemShut {NoStop}%
\bibitem [{\citenamefont {Miao}\ \emph {et~al.}(2021)\citenamefont {Miao},
  \citenamefont {Li}, \citenamefont {Meier}, \citenamefont {Huon},
  \citenamefont {Lee}, \citenamefont {Said}, \citenamefont {Lei}, \citenamefont
  {Ortiz}, \citenamefont {Wilson}, \citenamefont {Yin} \emph
  {et~al.}}]{Miao2021}%
  \BibitemOpen
  \bibfield  {author} {\bibinfo {author} {\bibfnamefont {H.}~\bibnamefont
  {Miao}}, \bibinfo {author} {\bibfnamefont {H.~X.}\ \bibnamefont {Li}},
  \bibinfo {author} {\bibfnamefont {W.}~\bibnamefont {Meier}}, \bibinfo
  {author} {\bibfnamefont {A.}~\bibnamefont {Huon}}, \bibinfo {author}
  {\bibfnamefont {H.~N.}\ \bibnamefont {Lee}}, \bibinfo {author} {\bibfnamefont
  {A.}~\bibnamefont {Said}}, \bibinfo {author} {\bibfnamefont {H.}~\bibnamefont
  {Lei}}, \bibinfo {author} {\bibfnamefont {B.}~\bibnamefont {Ortiz}}, \bibinfo
  {author} {\bibfnamefont {S.}~\bibnamefont {Wilson}}, \bibinfo {author}
  {\bibfnamefont {J.}~\bibnamefont {Yin}},  \emph {et~al.},\ }\bibfield
  {title} {\enquote {\bibinfo {title} {Geometry of the charge density wave in
  the kagome metal {AV}$_3${Sb}$_5$},}\ }\href
  {https://journals.aps.org/prb/abstract/10.1103/PhysRevB.104.195132}
  {\bibfield  {journal} {\bibinfo  {journal} {Phys. Rev. B}\ }\textbf {\bibinfo
  {volume} {104}},\ \bibinfo {pages} {195132} (\bibinfo {year}
  {2021})}\BibitemShut {NoStop}%
\bibitem [{\citenamefont {Ni}\ \emph {et~al.}(2021)\citenamefont {Ni} \emph
  {et~al.}}]{Ni2021}%
  \BibitemOpen
  \bibfield  {author} {\bibinfo {author} {\bibfnamefont {S.}~\bibnamefont {Ni}}
  \emph {et~al.},\ }\bibfield  {title} {\enquote {\bibinfo {title} {Anisotropic
  superconducting properties of kagome metal {CsV}$_3${Sb}$_5$},}\ }\href
  {https://iopscience.iop.org/article/10.1088/0256-307X/38/5/057403} {\bibfield
   {journal} {\bibinfo  {journal} {Chinese Physics Letters}\ }\textbf {\bibinfo
  {volume} {38}},\ \bibinfo {pages} {057403} (\bibinfo {year}
  {2021})}\BibitemShut {NoStop}%
\bibitem [{\citenamefont {Zhu}\ \emph {et~al.}(2022)\citenamefont {Zhu} \emph
  {et~al.}}]{Zhu2021}%
  \BibitemOpen
  \bibfield  {author} {\bibinfo {author} {\bibfnamefont {C.~C.}\ \bibnamefont
  {Zhu}} \emph {et~al.},\ }\bibfield  {title} {\enquote {\bibinfo {title}
  {Double-dome superconductivity under pressure in the {V}-based kagome metals
  {AV}$_3${Sb}$_5$ ({A}= {Rb and K})},}\ }\href
  {https://journals.aps.org/prb/abstract/10.1103/PhysRevB.105.094507}
  {\bibfield  {journal} {\bibinfo  {journal} {Phys. Rev. B}\ }\textbf {\bibinfo
  {volume} {105}},\ \bibinfo {pages} {094507} (\bibinfo {year}
  {2022})}\BibitemShut {NoStop}%
\bibitem [{\citenamefont {Chen}\ \emph {et~al.}(2021)\citenamefont {Chen} \emph
  {et~al.}}]{Chen2021b}%
  \BibitemOpen
  \bibfield  {author} {\bibinfo {author} {\bibfnamefont {K.~Y.}\ \bibnamefont
  {Chen}} \emph {et~al.},\ }\bibfield  {title} {\enquote {\bibinfo {title}
  {Double superconducting dome and triple enhancement of ${T_c}$ in the kagome
  superconductor {CsV}$_3${Sb}$_5$ under high pressure},}\ }\href
  {https://journals.aps.org/prl/abstract/10.1103/PhysRevLett.126.247001}
  {\bibfield  {journal} {\bibinfo  {journal} {Phys. Rev. Lett.}\ }\textbf
  {\bibinfo {volume} {126}},\ \bibinfo {pages} {247001} (\bibinfo {year}
  {2021})}\BibitemShut {NoStop}%
\bibitem [{\citenamefont {Du}\ \emph {et~al.}(2021)\citenamefont {Du},
  \citenamefont {Luo}, \citenamefont {Ortiz}, \citenamefont {Chen},
  \citenamefont {Duan}, \citenamefont {Zhang}, \citenamefont {Lu},
  \citenamefont {Wilson}, \citenamefont {Song},\ and\ \citenamefont
  {Yuan}}]{Du2021b}%
  \BibitemOpen
  \bibfield  {author} {\bibinfo {author} {\bibfnamefont {F.}~\bibnamefont
  {Du}}, \bibinfo {author} {\bibfnamefont {S.}~\bibnamefont {Luo}}, \bibinfo
  {author} {\bibfnamefont {B.~R.}\ \bibnamefont {Ortiz}}, \bibinfo {author}
  {\bibfnamefont {Y.}~\bibnamefont {Chen}}, \bibinfo {author} {\bibfnamefont
  {W.}~\bibnamefont {Duan}}, \bibinfo {author} {\bibfnamefont {D.}~\bibnamefont
  {Zhang}}, \bibinfo {author} {\bibfnamefont {X.}~\bibnamefont {Lu}}, \bibinfo
  {author} {\bibfnamefont {S.~D.}\ \bibnamefont {Wilson}}, \bibinfo {author}
  {\bibfnamefont {Y.}~\bibnamefont {Song}}, \ and\ \bibinfo {author}
  {\bibfnamefont {H.}~\bibnamefont {Yuan}},\ }\bibfield  {title} {\enquote
  {\bibinfo {title} {Pressure-induced double superconducting domes and charge
  instability in the kagome metal {KV}$_3${Sb}$_5$},}\ }\href
  {https://journals.aps.org/prb/abstract/10.1103/PhysRevB.103.L220504}
  {\bibfield  {journal} {\bibinfo  {journal} {Phys. Rev. B}\ }\textbf {\bibinfo
  {volume} {103}},\ \bibinfo {pages} {L220504} (\bibinfo {year}
  {2021})}\BibitemShut {NoStop}%
\bibitem [{\citenamefont {Zhang}\ \emph {et~al.}(2021)\citenamefont {Zhang},
  \citenamefont {Chen}, \citenamefont {Zhou}, \citenamefont {Yuan},
  \citenamefont {Wang}, \citenamefont {Wang}, \citenamefont {Yang},
  \citenamefont {An}, \citenamefont {Zhang}, \citenamefont {Zhu} \emph
  {et~al.}}]{Zhang2021}%
  \BibitemOpen
  \bibfield  {author} {\bibinfo {author} {\bibfnamefont {Z.}~\bibnamefont
  {Zhang}}, \bibinfo {author} {\bibfnamefont {Z.}~\bibnamefont {Chen}},
  \bibinfo {author} {\bibfnamefont {Y.}~\bibnamefont {Zhou}}, \bibinfo {author}
  {\bibfnamefont {Y.}~\bibnamefont {Yuan}}, \bibinfo {author} {\bibfnamefont
  {S.}~\bibnamefont {Wang}}, \bibinfo {author} {\bibfnamefont {J.}~\bibnamefont
  {Wang}}, \bibinfo {author} {\bibfnamefont {H.}~\bibnamefont {Yang}}, \bibinfo
  {author} {\bibfnamefont {C.}~\bibnamefont {An}}, \bibinfo {author}
  {\bibfnamefont {L.}~\bibnamefont {Zhang}}, \bibinfo {author} {\bibfnamefont
  {X.}~\bibnamefont {Zhu}},  \emph {et~al.},\ }\bibfield  {title} {\enquote
  {\bibinfo {title} {Pressure-induced reemergence of superconductivity in the
  topological kagome metal {CsV}$_3${Sb}$_5$},}\ }\href
  {https://journals.aps.org/prb/abstract/10.1103/PhysRevB.103.224513}
  {\bibfield  {journal} {\bibinfo  {journal} {Phys. Rev. B}\ }\textbf {\bibinfo
  {volume} {103}},\ \bibinfo {pages} {224513} (\bibinfo {year}
  {2021})}\BibitemShut {NoStop}%
\bibitem [{\citenamefont {Liang}\ \emph {et~al.}(2021)\citenamefont {Liang}
  \emph {et~al.}}]{Liang2021}%
  \BibitemOpen
  \bibfield  {author} {\bibinfo {author} {\bibfnamefont {Z.}~\bibnamefont
  {Liang}} \emph {et~al.},\ }\bibfield  {title} {\enquote {\bibinfo {title}
  {Three-dimensional charge density wave and surface-dependent vortex-core
  states in a kagome superconductor {CsV}$_3${Sb}$_5$},}\ }\href
  {https://journals.aps.org/prx/abstract/10.1103/PhysRevX.11.031026} {\bibfield
   {journal} {\bibinfo  {journal} {Physical Review X}\ }\textbf {\bibinfo
  {volume} {11}},\ \bibinfo {pages} {031026} (\bibinfo {year}
  {2021})}\BibitemShut {NoStop}%
\bibitem [{\citenamefont {Ortiz}\ \emph {et~al.}(2021)\citenamefont {Ortiz},
  \citenamefont {Teicher}, \citenamefont {Kautzsch}, \citenamefont {Sarte},
  \citenamefont {Ratcliff}, \citenamefont {Harter}, \citenamefont {Ruff},
  \citenamefont {Seshadri},\ and\ \citenamefont {Wilson}}]{ortiz2021fermi}%
  \BibitemOpen
  \bibfield  {author} {\bibinfo {author} {\bibfnamefont {B.~R.}\ \bibnamefont
  {Ortiz}}, \bibinfo {author} {\bibfnamefont {S.~M.}\ \bibnamefont {Teicher}},
  \bibinfo {author} {\bibfnamefont {L.}~\bibnamefont {Kautzsch}}, \bibinfo
  {author} {\bibfnamefont {P.~M.}\ \bibnamefont {Sarte}}, \bibinfo {author}
  {\bibfnamefont {N.}~\bibnamefont {Ratcliff}}, \bibinfo {author}
  {\bibfnamefont {J.}~\bibnamefont {Harter}}, \bibinfo {author} {\bibfnamefont
  {J.~P.}\ \bibnamefont {Ruff}}, \bibinfo {author} {\bibfnamefont
  {R.}~\bibnamefont {Seshadri}}, \ and\ \bibinfo {author} {\bibfnamefont
  {S.~D.}\ \bibnamefont {Wilson}},\ }\bibfield  {title} {\enquote {\bibinfo
  {title} {Fermi surface mapping and the nature of charge-density-wave order in
  the kagome superconductor {CsV}$_3${Sb}$_5$},}\ }\href
  {https://journals.aps.org/prx/abstract/10.1103/PhysRevX.11.041030} {\bibfield
   {journal} {\bibinfo  {journal} {Physical Review X}\ }\textbf {\bibinfo
  {volume} {11}},\ \bibinfo {pages} {041030} (\bibinfo {year}
  {2021})}\BibitemShut {NoStop}%
\bibitem [{\citenamefont {Oey}\ \emph {et~al.}(2022)\citenamefont {Oey},
  \citenamefont {Ortiz}, \citenamefont {Kaboudvand}, \citenamefont
  {Frassineti}, \citenamefont {Garcia}, \citenamefont {Cong}, \citenamefont
  {Sanna}, \citenamefont {Mitrovi{\'c}}, \citenamefont {Seshadri},\ and\
  \citenamefont {Wilson}}]{oey2022fermi}%
  \BibitemOpen
  \bibfield  {author} {\bibinfo {author} {\bibfnamefont {Y.~M.}\ \bibnamefont
  {Oey}}, \bibinfo {author} {\bibfnamefont {B.~R.}\ \bibnamefont {Ortiz}},
  \bibinfo {author} {\bibfnamefont {F.}~\bibnamefont {Kaboudvand}}, \bibinfo
  {author} {\bibfnamefont {J.}~\bibnamefont {Frassineti}}, \bibinfo {author}
  {\bibfnamefont {E.}~\bibnamefont {Garcia}}, \bibinfo {author} {\bibfnamefont
  {R.}~\bibnamefont {Cong}}, \bibinfo {author} {\bibfnamefont {S.}~\bibnamefont
  {Sanna}}, \bibinfo {author} {\bibfnamefont {V.~F.}\ \bibnamefont
  {Mitrovi{\'c}}}, \bibinfo {author} {\bibfnamefont {R.}~\bibnamefont
  {Seshadri}}, \ and\ \bibinfo {author} {\bibfnamefont {S.~D.}\ \bibnamefont
  {Wilson}},\ }\bibfield  {title} {\enquote {\bibinfo {title} {Fermi level
  tuning and double-dome superconductivity in the kagome metal
  {CsV}$_3${Sb}$_{5- x}${Sn}$_x$},}\ }\href
  {https://journals.aps.org/prmaterials/abstract/10.1103/PhysRevMaterials.6.L041801}
  {\bibfield  {journal} {\bibinfo  {journal} {Physical Review Materials}\
  }\textbf {\bibinfo {volume} {6}},\ \bibinfo {pages} {L041801} (\bibinfo
  {year} {2022})}\BibitemShut {NoStop}%
\bibitem [{\citenamefont {Kang}\ \emph {et~al.}(2023)\citenamefont {Kang},
  \citenamefont {Fang}, \citenamefont {Yoo}, \citenamefont {Ortiz},
  \citenamefont {Oey}, \citenamefont {Choi}, \citenamefont {Ryu}, \citenamefont
  {Kim}, \citenamefont {Jozwiak}, \citenamefont {Bostwick} \emph
  {et~al.}}]{kang2023charge}%
  \BibitemOpen
  \bibfield  {author} {\bibinfo {author} {\bibfnamefont {M.}~\bibnamefont
  {Kang}}, \bibinfo {author} {\bibfnamefont {S.}~\bibnamefont {Fang}}, \bibinfo
  {author} {\bibfnamefont {J.}~\bibnamefont {Yoo}}, \bibinfo {author}
  {\bibfnamefont {B.~R.}\ \bibnamefont {Ortiz}}, \bibinfo {author}
  {\bibfnamefont {Y.~M.}\ \bibnamefont {Oey}}, \bibinfo {author} {\bibfnamefont
  {J.}~\bibnamefont {Choi}}, \bibinfo {author} {\bibfnamefont {S.~H.}\
  \bibnamefont {Ryu}}, \bibinfo {author} {\bibfnamefont {J.}~\bibnamefont
  {Kim}}, \bibinfo {author} {\bibfnamefont {C.}~\bibnamefont {Jozwiak}},
  \bibinfo {author} {\bibfnamefont {A.}~\bibnamefont {Bostwick}},  \emph
  {et~al.},\ }\bibfield  {title} {\enquote {\bibinfo {title} {Charge order
  landscape and competition with superconductivity in kagome metals},}\ }\href
  {https://www.nature.com/articles/s41563-022-01375-2} {\bibfield  {journal}
  {\bibinfo  {journal} {Nat. Mater.}\ }\textbf {\bibinfo {volume} {22}},\
  \bibinfo {pages} {186} (\bibinfo {year} {2023})}\BibitemShut {NoStop}%
\bibitem [{\citenamefont {Huang}\ \emph {et~al.}(2025)\citenamefont {Huang},
  \citenamefont {Chen}, \citenamefont {Tan}, \citenamefont {Han}, \citenamefont
  {Ye}, \citenamefont {Hu}, \citenamefont {Zhao}, \citenamefont {Shen},
  \citenamefont {Yang}, \citenamefont {Yan} \emph
  {et~al.}}]{huang2025revealing}%
  \BibitemOpen
  \bibfield  {author} {\bibinfo {author} {\bibfnamefont {Z.}~\bibnamefont
  {Huang}}, \bibinfo {author} {\bibfnamefont {H.}~\bibnamefont {Chen}},
  \bibinfo {author} {\bibfnamefont {H.}~\bibnamefont {Tan}}, \bibinfo {author}
  {\bibfnamefont {X.}~\bibnamefont {Han}}, \bibinfo {author} {\bibfnamefont
  {Y.}~\bibnamefont {Ye}}, \bibinfo {author} {\bibfnamefont {B.}~\bibnamefont
  {Hu}}, \bibinfo {author} {\bibfnamefont {Z.}~\bibnamefont {Zhao}}, \bibinfo
  {author} {\bibfnamefont {C.}~\bibnamefont {Shen}}, \bibinfo {author}
  {\bibfnamefont {H.}~\bibnamefont {Yang}}, \bibinfo {author} {\bibfnamefont
  {B.}~\bibnamefont {Yan}},  \emph {et~al.},\ }\bibfield  {title} {\enquote
  {\bibinfo {title} {Revealing the orbital origins of exotic electronic states
  with ti substitution in kagome superconductor {CsV}$_3${Sb}$_5$},}\ }\href
  {https://journals.aps.org/prl/abstract/10.1103/PhysRevLett.134.056001}
  {\bibfield  {journal} {\bibinfo  {journal} {Phys. Rev. Lett.}\ }\textbf
  {\bibinfo {volume} {134}},\ \bibinfo {pages} {056001} (\bibinfo {year}
  {2025})}\BibitemShut {NoStop}%
\bibitem [{\citenamefont {Hossain}\ \emph {et~al.}(2025)\citenamefont
  {Hossain}, \citenamefont {Zhang}, \citenamefont {Ingham}, \citenamefont
  {Liu}, \citenamefont {Shao}, \citenamefont {Li}, \citenamefont {Wang},
  \citenamefont {Pokharel}, \citenamefont {Cheng}, \citenamefont {Jiang} \emph
  {et~al.}}]{hossain2025field}%
  \BibitemOpen
  \bibfield  {author} {\bibinfo {author} {\bibfnamefont {M.~S.}\ \bibnamefont
  {Hossain}}, \bibinfo {author} {\bibfnamefont {Q.}~\bibnamefont {Zhang}},
  \bibinfo {author} {\bibfnamefont {J.}~\bibnamefont {Ingham}}, \bibinfo
  {author} {\bibfnamefont {J.}~\bibnamefont {Liu}}, \bibinfo {author}
  {\bibfnamefont {S.}~\bibnamefont {Shao}}, \bibinfo {author} {\bibfnamefont
  {Y.}~\bibnamefont {Li}}, \bibinfo {author} {\bibfnamefont {Y.}~\bibnamefont
  {Wang}}, \bibinfo {author} {\bibfnamefont {B.~K.}\ \bibnamefont {Pokharel}},
  \bibinfo {author} {\bibfnamefont {Z.-J.}\ \bibnamefont {Cheng}}, \bibinfo
  {author} {\bibfnamefont {Y.-X.}\ \bibnamefont {Jiang}},  \emph {et~al.},\
  }\bibfield  {title} {\enquote {\bibinfo {title} {Field induced density wave
  in a kagome superconductor},}\ }\href {https://arxiv.org/abs/2501.13260}
  {\bibfield  {journal} {\bibinfo  {journal} {arXiv preprint arXiv:2501.13260}\
  } (\bibinfo {year} {2025})}\BibitemShut {NoStop}%
\bibitem [{\citenamefont {Park}\ \emph {et~al.}(2021)\citenamefont {Park},
  \citenamefont {Ye},\ and\ \citenamefont {Balents}}]{ParkPRB2021}%
  \BibitemOpen
  \bibfield  {author} {\bibinfo {author} {\bibfnamefont {T.}~\bibnamefont
  {Park}}, \bibinfo {author} {\bibfnamefont {M.}~\bibnamefont {Ye}}, \ and\
  \bibinfo {author} {\bibfnamefont {L.}~\bibnamefont {Balents}},\ }\bibfield
  {title} {\enquote {\bibinfo {title} {Electronic instabilities of kagome
  metals: Saddle points and landau theory},}\ }\href {\doibase
  10.1103/PhysRevB.104.035142} {\bibfield  {journal} {\bibinfo  {journal}
  {Phys. Rev. B}\ }\textbf {\bibinfo {volume} {104}},\ \bibinfo {pages}
  {035142} (\bibinfo {year} {2021})}\BibitemShut {NoStop}%
\bibitem [{\citenamefont {Christensen}\ \emph {et~al.}(2021)\citenamefont
  {Christensen}, \citenamefont {Birol}, \citenamefont {Andersen},\ and\
  \citenamefont {Fernandes}}]{christensen2021theory}%
  \BibitemOpen
  \bibfield  {author} {\bibinfo {author} {\bibfnamefont {M.~H.}\ \bibnamefont
  {Christensen}}, \bibinfo {author} {\bibfnamefont {T.}~\bibnamefont {Birol}},
  \bibinfo {author} {\bibfnamefont {B.~M.}\ \bibnamefont {Andersen}}, \ and\
  \bibinfo {author} {\bibfnamefont {R.~M.}\ \bibnamefont {Fernandes}},\
  }\bibfield  {title} {\enquote {\bibinfo {title} {Theory of the charge density
  wave in {AV}$_3${Sb}$_5$ kagome metals},}\ }\href
  {https://journals.aps.org/prb/abstract/10.1103/PhysRevB.104.214513}
  {\bibfield  {journal} {\bibinfo  {journal} {Phys. Rev. B}\ }\textbf {\bibinfo
  {volume} {104}},\ \bibinfo {pages} {214513} (\bibinfo {year}
  {2021})}\BibitemShut {NoStop}%
\bibitem [{\citenamefont {Wagner}\ \emph {et~al.}(2023)\citenamefont {Wagner},
  \citenamefont {Guo}, \citenamefont {Moll}, \citenamefont {Neupert},\ and\
  \citenamefont {Fischer}}]{WagnerPRB2023}%
  \BibitemOpen
  \bibfield  {author} {\bibinfo {author} {\bibfnamefont {G.}~\bibnamefont
  {Wagner}}, \bibinfo {author} {\bibfnamefont {C.}~\bibnamefont {Guo}},
  \bibinfo {author} {\bibfnamefont {P.~J.~W.}\ \bibnamefont {Moll}}, \bibinfo
  {author} {\bibfnamefont {T.}~\bibnamefont {Neupert}}, \ and\ \bibinfo
  {author} {\bibfnamefont {M.~H.}\ \bibnamefont {Fischer}},\ }\bibfield
  {title} {\enquote {\bibinfo {title} {Phenomenology of bond and flux orders in
  kagome metals},}\ }\href {\doibase 10.1103/PhysRevB.108.125136} {\bibfield
  {journal} {\bibinfo  {journal} {Phys. Rev. B}\ }\textbf {\bibinfo {volume}
  {108}},\ \bibinfo {pages} {125136} (\bibinfo {year} {2023})}\BibitemShut
  {NoStop}%
\bibitem [{\citenamefont {Asaba}\ \emph {et~al.}(2024)\citenamefont {Asaba},
  \citenamefont {Onishi}, \citenamefont {Kageyama}, \citenamefont {Kiyosue},
  \citenamefont {Ohtsuka}, \citenamefont {Suetsugu}, \citenamefont {Kohsaka},
  \citenamefont {Gaggl}, \citenamefont {Kasahara}, \citenamefont {Murayama},
  \citenamefont {Hashimoto}, \citenamefont {Tazai}, \citenamefont {Kontani},
  \citenamefont {Ortiz}, \citenamefont {Wilson}, \citenamefont {Li},
  \citenamefont {Wen}, \citenamefont {Shibauchi},\ and\ \citenamefont
  {Matsuda}}]{Asaba2024}%
  \BibitemOpen
  \bibfield  {author} {\bibinfo {author} {\bibfnamefont {T.}~\bibnamefont
  {Asaba}}, \bibinfo {author} {\bibfnamefont {A.}~\bibnamefont {Onishi}},
  \bibinfo {author} {\bibfnamefont {Y.}~\bibnamefont {Kageyama}}, \bibinfo
  {author} {\bibfnamefont {T.}~\bibnamefont {Kiyosue}}, \bibinfo {author}
  {\bibfnamefont {K.}~\bibnamefont {Ohtsuka}}, \bibinfo {author} {\bibfnamefont
  {S.}~\bibnamefont {Suetsugu}}, \bibinfo {author} {\bibfnamefont
  {Y.}~\bibnamefont {Kohsaka}}, \bibinfo {author} {\bibfnamefont
  {T.}~\bibnamefont {Gaggl}}, \bibinfo {author} {\bibfnamefont
  {Y.}~\bibnamefont {Kasahara}}, \bibinfo {author} {\bibfnamefont
  {H.}~\bibnamefont {Murayama}}, \bibinfo {author} {\bibfnamefont
  {K.}~\bibnamefont {Hashimoto}}, \bibinfo {author} {\bibfnamefont
  {R.}~\bibnamefont {Tazai}}, \bibinfo {author} {\bibfnamefont
  {H.}~\bibnamefont {Kontani}}, \bibinfo {author} {\bibfnamefont {B.~R.}\
  \bibnamefont {Ortiz}}, \bibinfo {author} {\bibfnamefont {S.~D.}\ \bibnamefont
  {Wilson}}, \bibinfo {author} {\bibfnamefont {Q.}~\bibnamefont {Li}}, \bibinfo
  {author} {\bibfnamefont {H.-H.}\ \bibnamefont {Wen}}, \bibinfo {author}
  {\bibfnamefont {T.}~\bibnamefont {Shibauchi}}, \ and\ \bibinfo {author}
  {\bibfnamefont {Y.}~\bibnamefont {Matsuda}},\ }\bibfield  {title} {\enquote
  {\bibinfo {title} {Evidence for an odd-parity nematic phase above the
  charge-density-wave transition in a kagome metal},}\ }\href {\doibase
  10.1038/s41567-023-02272-4} {\bibfield  {journal} {\bibinfo  {journal}
  {Nature Physics}\ }\textbf {\bibinfo {volume} {20}},\ \bibinfo {pages} {40}
  (\bibinfo {year} {2024})}\BibitemShut {NoStop}%
\bibitem [{\citenamefont {Guo}\ \emph {et~al.}(2024)\citenamefont {Guo},
  \citenamefont {van Delft}, \citenamefont {Gutierrez-Amigo}, \citenamefont
  {Chen}, \citenamefont {Putzke}, \citenamefont {Wagner}, \citenamefont
  {Fischer}, \citenamefont {Neupert}, \citenamefont {Errea}, \citenamefont
  {Vergniory}, \citenamefont {Wiedmann}, \citenamefont {Felser},\ and\
  \citenamefont {Moll}}]{emcha_Guo2024}%
  \BibitemOpen
  \bibfield  {author} {\bibinfo {author} {\bibfnamefont {C.}~\bibnamefont
  {Guo}}, \bibinfo {author} {\bibfnamefont {M.~R.}\ \bibnamefont {van Delft}},
  \bibinfo {author} {\bibfnamefont {M.}~\bibnamefont {Gutierrez-Amigo}},
  \bibinfo {author} {\bibfnamefont {D.}~\bibnamefont {Chen}}, \bibinfo {author}
  {\bibfnamefont {C.}~\bibnamefont {Putzke}}, \bibinfo {author} {\bibfnamefont
  {G.}~\bibnamefont {Wagner}}, \bibinfo {author} {\bibfnamefont {M.~H.}\
  \bibnamefont {Fischer}}, \bibinfo {author} {\bibfnamefont {T.}~\bibnamefont
  {Neupert}}, \bibinfo {author} {\bibfnamefont {I.}~\bibnamefont {Errea}},
  \bibinfo {author} {\bibfnamefont {M.~G.}\ \bibnamefont {Vergniory}}, \bibinfo
  {author} {\bibfnamefont {S.}~\bibnamefont {Wiedmann}}, \bibinfo {author}
  {\bibfnamefont {C.}~\bibnamefont {Felser}}, \ and\ \bibinfo {author}
  {\bibfnamefont {P.~J.~W.}\ \bibnamefont {Moll}},\ }\bibfield  {title}
  {\enquote {\bibinfo {title} {Distinct switching of chiral transport in the
  kagome metals {KV}$_3${Sb}$_5$ and {CsV}$_3${Sb}$_5$},}\ }\href
  {https://doi.org/10.1038/s41535-024-00629-3} {\bibfield  {journal} {\bibinfo
  {journal} {npj Quantum Materials}\ }\textbf {\bibinfo {volume} {9}},\
  \bibinfo {pages} {20} (\bibinfo {year} {2024})}\BibitemShut {NoStop}%
\bibitem [{\citenamefont {Kang}\ \emph {et~al.}(2022)\citenamefont {Kang},
  \citenamefont {Fang}, \citenamefont {Kim}, \citenamefont {Ortiz},
  \citenamefont {Ryu}, \citenamefont {Kim}, \citenamefont {Yoo}, \citenamefont
  {Sangiovanni}, \citenamefont {Di~Sante}, \citenamefont {Park}, \citenamefont
  {Jozwiak}, \citenamefont {Bostwick}, \citenamefont {Rotenberg}, \citenamefont
  {Kaxiras}, \citenamefont {Wilson}, \citenamefont {Park},\ and\ \citenamefont
  {Comin}}]{Kang2022}%
  \BibitemOpen
  \bibfield  {author} {\bibinfo {author} {\bibfnamefont {M.}~\bibnamefont
  {Kang}}, \bibinfo {author} {\bibfnamefont {S.}~\bibnamefont {Fang}}, \bibinfo
  {author} {\bibfnamefont {J.-K.}\ \bibnamefont {Kim}}, \bibinfo {author}
  {\bibfnamefont {B.~R.}\ \bibnamefont {Ortiz}}, \bibinfo {author}
  {\bibfnamefont {S.~H.}\ \bibnamefont {Ryu}}, \bibinfo {author} {\bibfnamefont
  {J.}~\bibnamefont {Kim}}, \bibinfo {author} {\bibfnamefont {J.}~\bibnamefont
  {Yoo}}, \bibinfo {author} {\bibfnamefont {G.}~\bibnamefont {Sangiovanni}},
  \bibinfo {author} {\bibfnamefont {D.}~\bibnamefont {Di~Sante}}, \bibinfo
  {author} {\bibfnamefont {B.-G.}\ \bibnamefont {Park}}, \bibinfo {author}
  {\bibfnamefont {C.}~\bibnamefont {Jozwiak}}, \bibinfo {author} {\bibfnamefont
  {A.}~\bibnamefont {Bostwick}}, \bibinfo {author} {\bibfnamefont
  {E.}~\bibnamefont {Rotenberg}}, \bibinfo {author} {\bibfnamefont
  {E.}~\bibnamefont {Kaxiras}}, \bibinfo {author} {\bibfnamefont {S.~D.}\
  \bibnamefont {Wilson}}, \bibinfo {author} {\bibfnamefont {J.-H.}\
  \bibnamefont {Park}}, \ and\ \bibinfo {author} {\bibfnamefont
  {R.}~\bibnamefont {Comin}},\ }\bibfield  {title} {\enquote {\bibinfo {title}
  {Twofold van {Hove} singularity and origin of charge order in topological
  kagome superconductor {CsV}$_3${Sb}$_5$},}\ }\href {\doibase
  10.1038/s41567-021-01451-5} {\bibfield  {journal} {\bibinfo  {journal} {Nat.
  Phys.}\ }\textbf {\bibinfo {volume} {18}},\ \bibinfo {pages} {301} (\bibinfo
  {year} {2022})}\BibitemShut {NoStop}%
\bibitem [{\citenamefont {Hu}\ \emph {et~al.}(2022)\citenamefont {Hu},
  \citenamefont {Wu}, \citenamefont {Ortiz}, \citenamefont {Ju}, \citenamefont
  {Han}, \citenamefont {Ma}, \citenamefont {Plumb}, \citenamefont {Radovic},
  \citenamefont {Thomale}, \citenamefont {Wilson}, \citenamefont {Schnyder},\
  and\ \citenamefont {Shi}}]{Hu2022}%
  \BibitemOpen
  \bibfield  {author} {\bibinfo {author} {\bibfnamefont {Y.}~\bibnamefont
  {Hu}}, \bibinfo {author} {\bibfnamefont {X.}~\bibnamefont {Wu}}, \bibinfo
  {author} {\bibfnamefont {B.~R.}\ \bibnamefont {Ortiz}}, \bibinfo {author}
  {\bibfnamefont {S.}~\bibnamefont {Ju}}, \bibinfo {author} {\bibfnamefont
  {X.}~\bibnamefont {Han}}, \bibinfo {author} {\bibfnamefont {J.}~\bibnamefont
  {Ma}}, \bibinfo {author} {\bibfnamefont {N.~C.}\ \bibnamefont {Plumb}},
  \bibinfo {author} {\bibfnamefont {M.}~\bibnamefont {Radovic}}, \bibinfo
  {author} {\bibfnamefont {R.}~\bibnamefont {Thomale}}, \bibinfo {author}
  {\bibfnamefont {S.~D.}\ \bibnamefont {Wilson}}, \bibinfo {author}
  {\bibfnamefont {A.~P.}\ \bibnamefont {Schnyder}}, \ and\ \bibinfo {author}
  {\bibfnamefont {M.}~\bibnamefont {Shi}},\ }\bibfield  {title} {\enquote
  {\bibinfo {title} {Rich nature of van {Hove} singularities in kagome
  superconductor {CsV}$_3${Sb}$_5$},}\ }\href {\doibase
  10.1038/s41467-022-29828-x} {\bibfield  {journal} {\bibinfo  {journal}
  {Nature Communications}\ }\textbf {\bibinfo {volume} {13}},\ \bibinfo {pages}
  {2220} (\bibinfo {year} {2022})}\BibitemShut {NoStop}%
\bibitem [{\citenamefont {Wu}\ \emph {et~al.}(2021)\citenamefont {Wu},
  \citenamefont {Schwemmer}, \citenamefont {M\"uller}, \citenamefont
  {Consiglio}, \citenamefont {Sangiovanni}, \citenamefont {Di~Sante},
  \citenamefont {Iqbal}, \citenamefont {Hanke}, \citenamefont {Schnyder},
  \citenamefont {Denner}, \citenamefont {Fischer}, \citenamefont {Neupert},\
  and\ \citenamefont {Thomale}}]{Thomale_PRL2021}%
  \BibitemOpen
  \bibfield  {author} {\bibinfo {author} {\bibfnamefont {X.}~\bibnamefont
  {Wu}}, \bibinfo {author} {\bibfnamefont {T.}~\bibnamefont {Schwemmer}},
  \bibinfo {author} {\bibfnamefont {T.}~\bibnamefont {M\"uller}}, \bibinfo
  {author} {\bibfnamefont {A.}~\bibnamefont {Consiglio}}, \bibinfo {author}
  {\bibfnamefont {G.}~\bibnamefont {Sangiovanni}}, \bibinfo {author}
  {\bibfnamefont {D.}~\bibnamefont {Di~Sante}}, \bibinfo {author}
  {\bibfnamefont {Y.}~\bibnamefont {Iqbal}}, \bibinfo {author} {\bibfnamefont
  {W.}~\bibnamefont {Hanke}}, \bibinfo {author} {\bibfnamefont {A.~P.}\
  \bibnamefont {Schnyder}}, \bibinfo {author} {\bibfnamefont {M.~M.}\
  \bibnamefont {Denner}}, \bibinfo {author} {\bibfnamefont {M.~H.}\
  \bibnamefont {Fischer}}, \bibinfo {author} {\bibfnamefont {T.}~\bibnamefont
  {Neupert}}, \ and\ \bibinfo {author} {\bibfnamefont {R.}~\bibnamefont
  {Thomale}},\ }\bibfield  {title} {\enquote {\bibinfo {title} {Nature of
  unconventional pairing in the kagome superconductors ${A}${V}$_3${Sb}$_5$
  (${A}$={K,Kb,Cs})},}\ }\href {\doibase 10.1103/PhysRevLett.127.177001}
  {\bibfield  {journal} {\bibinfo  {journal} {Phys. Rev. Lett.}\ }\textbf
  {\bibinfo {volume} {127}},\ \bibinfo {pages} {177001} (\bibinfo {year}
  {2021})}\BibitemShut {NoStop}%
\bibitem [{\citenamefont {Scammell}\ \emph {et~al.}(2023)\citenamefont
  {Scammell}, \citenamefont {Ingham}, \citenamefont {Li},\ and\ \citenamefont
  {Sushkov}}]{Scammell2023}%
  \BibitemOpen
  \bibfield  {author} {\bibinfo {author} {\bibfnamefont {H.~D.}\ \bibnamefont
  {Scammell}}, \bibinfo {author} {\bibfnamefont {J.}~\bibnamefont {Ingham}},
  \bibinfo {author} {\bibfnamefont {T.}~\bibnamefont {Li}}, \ and\ \bibinfo
  {author} {\bibfnamefont {O.~P.}\ \bibnamefont {Sushkov}},\ }\bibfield
  {title} {\enquote {\bibinfo {title} {Chiral excitonic order from twofold van
  hove singularities in kagome metals},}\ }\href
  {https://www.nature.com/articles/s41467-023-35987-2} {\bibfield  {journal}
  {\bibinfo  {journal} {Nature Communications}\ }\textbf {\bibinfo {volume}
  {14}},\ \bibinfo {pages} {605} (\bibinfo {year} {2023})}\BibitemShut
  {NoStop}%
\bibitem [{\citenamefont {Xing}\ \emph {et~al.}(2024)\citenamefont {Xing},
  \citenamefont {Bae}, \citenamefont {Ritz}, \citenamefont {Yang},
  \citenamefont {Birol}, \citenamefont {Capa~Salinas}, \citenamefont {Ortiz},
  \citenamefont {Wilson}, \citenamefont {Wang}, \citenamefont {Fernandes},\
  and\ \citenamefont {Madhavan}}]{Xing2024}%
  \BibitemOpen
  \bibfield  {author} {\bibinfo {author} {\bibfnamefont {Y.}~\bibnamefont
  {Xing}}, \bibinfo {author} {\bibfnamefont {S.}~\bibnamefont {Bae}}, \bibinfo
  {author} {\bibfnamefont {E.}~\bibnamefont {Ritz}}, \bibinfo {author}
  {\bibfnamefont {F.}~\bibnamefont {Yang}}, \bibinfo {author} {\bibfnamefont
  {T.}~\bibnamefont {Birol}}, \bibinfo {author} {\bibfnamefont {A.~N.}\
  \bibnamefont {Capa~Salinas}}, \bibinfo {author} {\bibfnamefont {B.~R.}\
  \bibnamefont {Ortiz}}, \bibinfo {author} {\bibfnamefont {S.~D.}\ \bibnamefont
  {Wilson}}, \bibinfo {author} {\bibfnamefont {Z.}~\bibnamefont {Wang}},
  \bibinfo {author} {\bibfnamefont {R.~M.}\ \bibnamefont {Fernandes}}, \ and\
  \bibinfo {author} {\bibfnamefont {V.}~\bibnamefont {Madhavan}},\ }\bibfield
  {title} {\enquote {\bibinfo {title} {Optical manipulation of the
  charge-density-wave state in rbv3sb5},}\ }\href {\doibase
  10.1038/s41586-024-07519-5} {\bibfield  {journal} {\bibinfo  {journal}
  {Nature}\ }\textbf {\bibinfo {volume} {631}},\ \bibinfo {pages} {60}
  (\bibinfo {year} {2024})}\BibitemShut {NoStop}%
\bibitem [{\citenamefont {Nandkishore}\ \emph {et~al.}(2012)\citenamefont
  {Nandkishore}, \citenamefont {Chern},\ and\ \citenamefont
  {Chubukov}}]{NandkishoreChubukovPRL2012}%
  \BibitemOpen
  \bibfield  {author} {\bibinfo {author} {\bibfnamefont {R.}~\bibnamefont
  {Nandkishore}}, \bibinfo {author} {\bibfnamefont {G.-W.}\ \bibnamefont
  {Chern}}, \ and\ \bibinfo {author} {\bibfnamefont {A.~V.}\ \bibnamefont
  {Chubukov}},\ }\bibfield  {title} {\enquote {\bibinfo {title} {Itinerant
  half-metal spin-density-wave state on the hexagonal lattice},}\ }\href
  {\doibase 10.1103/PhysRevLett.108.227204} {\bibfield  {journal} {\bibinfo
  {journal} {Phys. Rev. Lett.}\ }\textbf {\bibinfo {volume} {108}},\ \bibinfo
  {pages} {227204} (\bibinfo {year} {2012})}\BibitemShut {NoStop}%
\bibitem [{\citenamefont {Ortiz}\ \emph
  {et~al.}(2020{\natexlab{b}})\citenamefont {Ortiz}, \citenamefont {Teicher},
  \citenamefont {Hu}, \citenamefont {Zuo}, \citenamefont {Sarte}, \citenamefont
  {Schueller}, \citenamefont {Abeykoon}, \citenamefont {Krogstad},
  \citenamefont {Rosenkranz}, \citenamefont {Osborn}, \citenamefont {Seshadri},
  \citenamefont {Balents}, \citenamefont {He},\ and\ \citenamefont
  {Wilson}}]{OrtizTeicherPRL2020}%
  \BibitemOpen
  \bibfield  {author} {\bibinfo {author} {\bibfnamefont {B.~R.}\ \bibnamefont
  {Ortiz}}, \bibinfo {author} {\bibfnamefont {S.~M.~L.}\ \bibnamefont
  {Teicher}}, \bibinfo {author} {\bibfnamefont {Y.}~\bibnamefont {Hu}},
  \bibinfo {author} {\bibfnamefont {J.~L.}\ \bibnamefont {Zuo}}, \bibinfo
  {author} {\bibfnamefont {P.~M.}\ \bibnamefont {Sarte}}, \bibinfo {author}
  {\bibfnamefont {E.~C.}\ \bibnamefont {Schueller}}, \bibinfo {author}
  {\bibfnamefont {A.~M.~M.}\ \bibnamefont {Abeykoon}}, \bibinfo {author}
  {\bibfnamefont {M.~J.}\ \bibnamefont {Krogstad}}, \bibinfo {author}
  {\bibfnamefont {S.}~\bibnamefont {Rosenkranz}}, \bibinfo {author}
  {\bibfnamefont {R.}~\bibnamefont {Osborn}}, \bibinfo {author} {\bibfnamefont
  {R.}~\bibnamefont {Seshadri}}, \bibinfo {author} {\bibfnamefont
  {L.}~\bibnamefont {Balents}}, \bibinfo {author} {\bibfnamefont
  {J.}~\bibnamefont {He}}, \ and\ \bibinfo {author} {\bibfnamefont {S.~D.}\
  \bibnamefont {Wilson}},\ }\bibfield  {title} {\enquote {\bibinfo {title}
  {$\mathrm{Cs}{\mathrm{v}}_{3}{\mathrm{sb}}_{5}$: A ${\mathbb{z}}_{2}$
  topological kagome metal with a superconducting ground state},}\ }\href
  {\doibase 10.1103/PhysRevLett.125.247002} {\bibfield  {journal} {\bibinfo
  {journal} {Phys. Rev. Lett.}\ }\textbf {\bibinfo {volume} {125}},\ \bibinfo
  {pages} {247002} (\bibinfo {year} {2020}{\natexlab{b}})}\BibitemShut
  {NoStop}%
\end{thebibliography}
\end{document}